\documentclass[fleqn,usenatbib]{mnras}

\usepackage{newtxtext,newtxmath}

\usepackage[T1]{fontenc}

\DeclareRobustCommand{\VAN}[3]{#2}
\let\VANthebibliography\thebibliography
\def\thebibliography{\DeclareRobustCommand{\VAN}[3]{##3}\VANthebibliography}

\usepackage{graphicx}	
\usepackage{float}
\usepackage{stfloats} 
\usepackage{natbib}
\usepackage{amsmath}	
\usepackage{threeparttable}

\newcommand{\hto}{{\hbox {H$_{2}$O}}\,}
\def\htot#1#2#3#4#5#6{\hbox {\hto(\t#1#2#3#4#5#6)}}
\def\t#1#2#3#4#5#6{{\hbox {$#1_{#2#3}\text{--}#4_{#5#6}$}}}
\newcommand{\lhto}{$L_\mathrm{H{_2}O}$\,}
\newcommand{\ir}{$L_\mathrm{IR}$\,}

\title[H$_2$O in ULIRGs]{{\color{black}The first ground-based detection of the 
752 GHz water line in local ultra-luminous infrared galaxies using APEX-SEPIA}}
\author[D. Quinatoa et al.]
{Daysi Quinatoa,$^{1}$\thanks{E-mail: daysi.quinatoa@postgrado.uv.cl}
Chentao Yang,$^{2}$
Edo Ibar,$^{1}$
Elizabeth Humphreys,$^{3,4}$ \and
Susanne Aalto,$^{2}$ 
Loreto Barcos-Mu\~noz,$^{5,6}$ 
Eduardo Gonz\'alez-Alfonso,$^{7}$ 
Violette Impellizzeri,$^{8}$ \and
Yara Jaff\'e,$^{9,1}$ 
Lijie Liu,$^{10,11}$  
Sergio Mart\'in, $^{3,4}$ 
Axel Weiss,$^{12}$
Zhi-Yu Zhang$^{13,14}$
\\
$^{1}$Instituto de Física y Astronomía, Universidad de Valparaíso, Avda. Gran Bretaña 1111, Valparaíso, Chile\\
$^{2}$Department of Earth and Space Sciences, Chalmers University of Technology, Onsala Observatory, 439 94 Onsala, Sweden\\
$^{3}$Joint ALMA Observatory, Alonso de Córdova 3107 Vitacura, Santiago 763-0355, Chile
\\
$^{4}$European Southern Observatory, Alonso de Córdova, 3107, Vitacura, Santiago 763-0355, Chile\\
$^{5}$National Radio Astronomy Observatory, 520 Edgemont Road, Charlottesville, VA, 22903, USA\\
$^{6}$Department of Astronomy, University of Virginia, 530 McCormick Road, Charlottesville, VA, 22903, USA\\
$^{7}$Universidad de Alcalá, Departamento de Física y Matemáticas, Campus Universitario, 28871 Alcalá de Henares, Madrid, Spain\\
$^{8}$Leiden Observatory, Leiden University, PO Box 9513, 2300 RA, Leiden, The Netherlands\\
$^{9}$Departamento de Física, Universidad Técnica Federico Santa María
Casilla 110-V, Valparaíso, Chile\\
$^{10}$Cosmic Dawn Center (DAWN), Denmark\\
$^{11}$DTU-Space, Elektrovej, Building 328 , 2800, Kgs. Lyngby, Denmark\\
$^{12}$Max-Planck-Institut für Radioastronomie, Auf dem Hügel 69, D-53121 Bonn, Germany\\
$^{13}$ School of Astronomy and Space Science, Nanjing University, Nanjing 210023, People’s Republic of China\\
$^{14}$ Key Laboratory of Modern Astronomy and Astrophysics (Nanjing University), Ministry of Education, Nanjing 210023, People’s Republic of China\\
}

\date{Accepted 2023 November 03. Received 2023 October 30; in original form 2023 August 23}

\pubyear{2023}

\begin{document}

\label{firstpage}
\pagerange{\pageref{firstpage}--\pageref{lastpage}}
\maketitle

\begin{abstract}
We report the first ground-based detection of the water line p-\htot211202 
at 752.033\,GHz in three $z<0.08$ ultra-luminous infrared galaxies (ULIRGs):  
IRAS\,06035-7102, IRAS\,17207-0014 and IRAS\,09022-3615. Using the 
Atacama Pathfinder EXperiment (APEX), with its Swedish-ESO PI Instrument for APEX
(SEPIA) band-9 receiver, we detect 
this \hto\ line with overall signal-to-noise ratios of 8--10 in all three 
galaxies. Notably, this is the first detection of this line in IRAS\,06035-7102.
Our new APEX-measured fluxes, between 145 to 705 Jy\,km\,s$^{-1}$, 
are compared with previous values taken from {\it Herschel} SPIRE FTS. 
We highlight the great capabilities 
of APEX for resolving the \hto\ line profiles {\color{black}with high spectral resolutions} while also improving by a factor of two
the significance of the detection within moderate integration times. 
While exploring the correlation between the p-H$_2$O(2$_{11}$--2$_{02}$) 
and the total infrared luminosity, our galaxies are found to follow the trend
at the bright 
end of the local ULIRG's distribution. 
The p-H$_2$O(2$_{11}$--2$_{02}$) line spectra are compared to the mid-$J$ CO {\color{black} and HCN  
spectra, and dust continuum previously observed with ALMA.} 
In the complex interacting system IRAS\,09022-3615, {\color{black}the profile of the water emission line
is offset in velocity with respect} to the ALMA\,CO($J=4-3$) emission. 
For IRAS\,17207-0014 and IRAS\,06035-7102, the profiles between the water 
line and the CO lines are spectroscopically aligned. 
This pilot study demonstrates the feasibility of directly conducting 
ground-based high-frequency observations of this key water line, opening 
the possibility of detailed follow-up campaigns to tackle its nature.
\end{abstract}


\begin{keywords}
ISM: molecules -- galaxies: ISM -- infrared: galaxies
\end{keywords}


\section{Introduction}
Ultra-Luminous InfraRed Galaxies (ULIRGs; with total infrared luminosity \ir\ between 10$^{12}$\,$L_\odot$ and 10$^{13}$\,$L_\odot$) 
are part of the most extreme galaxy populations in the nearby Universe \citep[][]{Sanders2003}{}{}. 
They are characterized to be rich in dust and molecular gas 
\citep[e.g.][]{Solomon1997,genzel1998, Greve2005} and {\color{black} to have higher gas fractions 
compared }to normal star-forming galaxies \citep[][]{gao2004}. ULIRGs are mostly late-stage 
mergers \citep[e.g.][]{Sanders1996}{}{} understood as a 
transitional phase in the evolution of galaxies, passing through a brief starburst phase triggered 
by a major interaction that cause the in-falling of the interstellar medium (ISM) to the innermost 
regions, likely triggering nuclear starburst and active galactic nuclei (AGN) but with complex 
evolution links between the two \citep[e.g.][]{farrah2001,Lonsdale2006}. After the molecular gas is consumed, 
they will probably end up as passive galaxies hosting supermassive black holes in their centres. 
Therefore, ULIRGs offer unique insights into our understanding of galaxy evolution. 
To describe their nature, it is imperative to understand why they have such intense, compact 
star formation and what powers the nuclear star formation activity. The majority of the 
radiation produced in their nuclei, whether by an AGN or by star formation, is absorbed by 
dust and re-emitted in the far-infrared (far-IR). Because of the intense obscuration, identifying the predominant power 
source of local ULIRGs is challenging. Using emission lines from {\color{black}molecules such as OH and H$_2$O, 
whose excitation depend on the far-IR continuum photons, could be} a 
potential diagnostic to distinguish AGN from starburst activity 
\citep[][]{GonzalesAlfonso2010,vanderWerf2011,Pensabene22,decarli2023}. 

H$_2$O is the third most abundant molecule in the ISM, either in the gas phase in warm regions or in the solid phase on dust mantles \citep[e.g.][]{vanDishoeck2013}, and serves as one of the most important coolants of the cold molecular gas \citep{Neufeld}. 
{\color{black} In particular, the thermal para-H$_2$O(2$_{11}$--2$_{02}$) line is emitted at 752.033\,GHz rest-frame and has an upper energy level of $E_\mathrm{up}/k_\mathrm{B}=136.9$\,K. In Galactic star forming regions, this line is predominantly originated by shocks that trace high density and temperature gas} \citep[e.g.][]{Mottram2014,vanDishoeck2021}.
On the other hand, surveys of submillimeter (submm) \hto\ lines with {\it Herschel} in local galaxies revealed that this water line is one of the brightest submm \hto\ lines \citep{Yang2013,Lu2017}. The modelling of water in galaxies, assuming different phases of the ISM, with different dust temperatures, shows that this line comes predominantly from the warm phase with $T_\mathrm{dust}\sim 45\text{--}75$~K \citep{Gonzales2014}. The p-\htot211202\ line can be excited via collision in warm dense conditions with a column density of $N_\mathrm{H_2O}\sim(0.5-2)\times 10^{17}$\,cm$^{-2}$, sharing a common spatial distribution with those regions traced by mid-$J$ CO lines in star-forming galaxies \citep[][]{Liu2017, Gonzales2014}. 
Additionally, H$_2$O lines can be excited radiatively by far-IR photons. This so-called pumping mechanism of p-\htot211202 is induced by the 101$\mu$m continuum, which excites the H$_2$O molecules from the base energy level 1$_{11}$ to 2$_{20}$. This level 2$_{20}$ then cascades down to 2$_{11}$ and later to 2$_{02}$ producing line emission at 1229, 752, and 988\,GHz respectively. Furthermore,  the combination of far-IR 75\,$\mu$m absorption ortho-\htot321212\ and the \t321312\ 1163 GHz emission enhance the radiative excitation for the submm lines with $E_\mathrm{up}/k_\mathrm{B}<300$\,K, including the 752 GHz one \citep[][]{Gonzalez2022}. The p-\htot211202 far-IR pumping makes \hto\ lines a prominent tracer of the conditions of the far-IR field, especially in highly obscured regions of galaxies \cite[e.g.][]{Gonzales2014}. Previous studies have shown that the p-\htot211202\ is well correlated with {\color{black}the total infrared luminosity in both local and high-redshift galaxies} \citep[][]{Omont2013, Yang2013, Yang2016, Jarugula2019, Berta2023}. The nearly linear correlation {\color{black}between \lhto and \ir lines suggest that water} can be excited to high energy levels by far infrared pumping and provides another approach for tracing the far infrared field in star-forming galaxies. The correlation can be a natural consequence of far-IR pumping \hto\ excitation \citep{Gonzales2014}.

In dusty star-forming galaxies, H$_2$O lines trace both, the properties of the 
molecular gas and the dust content, offering a powerful tool to study their 
physical nature. 
This becomes especially important in ULIRGs as they 
present ideal conditions for water emission, intense far-IR radiation fields and warm dense gas content. In fact, the H$_2$O lines 
can offer unique insights into the innermost dust-obscured regions with very high opacity, providing essential information 
on the ISM physical conditions of the most extreme ULIRGs \citep[e.g.][]{Falstad2017, Liu2017, Yang2020, GA21}, where often the p-\htot211202\ line becomes optically thick \citep[e.g.][]{Gonzales2014}.

Although thermal H$_2$O lines provide powerful line diagnostics, their detection from ground-based observatories is impeded by the presence of water vapour in the Earth's atmosphere, which significantly reduces atmospheric transmission. From space, the H$_2$O lines have been studied in large samples of far-IR bright galaxies with limited spectral resolutions with {\it Herschel Space Observatory} SPIRE-FTS \citep[e.g.][]{Yang2013, Pearson2016}, despite a {\color{black}handful of sources studied with high-spectral resolution }with {\it Herschel} HIFI \citep{Liu2017}. {\color{black}It is known that the  p-\htot211202 line is ubiquitous in the spectra of molecular clouds in the Milky Way and local galaxies, but it has also been found to be a prominent emission line amongst the water lines detected in high-redshift infrared bright galaxies, both unlensed \citep[e.g.][]{2016A&A...591A..73G, 2017ApJ...850....1R, Casey_2019,lehnert2020,2020ApJ...889..162L,Stanley2021,Pensabene2021} and gravitationally amplified \citep[e.g.][]{Omont2011,Lis2011, Bradford2011, Combes2012, Omont2013, 2013Natur.496..329R, 2013ApJ...779...67B, Yang2016, Jarugula2019, Jarugula2021,  Apostolovski2019, 2019A&A...624A.138Y, 2019ApJ...880..153Y, Yang2020, 2021A&A...646A.122B}.
The similarity found in profiles between water and mid-$J$ to high-$J$ level CO transitions strongly suggests that water emission originates from regions of active star formation (e.g.\ see Figure~13 from \citealt{Yang2017}). Exploring the properties of water lines then becomes a key aspect for characterizing} the cosmic star formation history of the Universe.

It is important to highlight that most of the detected submm water lines from local galaxies have been made by {\it Herschel} (see \citealt{Weiss2010,Yang2013,Liu2017}),
nevertheless as this is {\color{black} no longer available, and until the next generation of far-infrared space telescopes, 
exploiting ground-based submm facilities are the only possibility to explore high-frequencies to measure water lines with high spectral resolution and, in the case of ALMA,
unprecedented angular resolution.} 

In this work, we present high-spectral resolution observations of the p-\htot211202\ {\color{black}
in three low-redshift ULIRGs }(IRAS\,06035-7102, IRAS\,17208-0014 and 
IRAS\,09022-3615) using the single-dish APEX telescope with the SEPIA receiver 
in Band-9 \citep{Baryshev2015,Belitsky2018}. Throughout this work, we assumed a 
$\Lambda$CDM cosmology with $\Omega_{\rm M}=0.3$, $\Omega_\Lambda=0.7$ and 
$H_0=75\,{\rm km\,s^{-1}\,Mpc^{-1}}$.

\section{Data}

\subsection{{\color{black}Source selection}}
{\color{black}The three sources were selected from a sample of ULIRGs observed in water lines with {\it Herschel} and presented in \citet[][]{Yang2013}. From this sample, we selected the sources below declination $+25^\circ$ and, in order to ensure a good atmospheric transmission, we excluded} sources whose observed water line p-\htot211202\ (752.033\,GHz rest-frame) frequencies are higher than 722\,GHz. Eight ULIRGs fulfil the selection criteria. Based on integration time estimates, considering expected line fluxes and atmospheric transmission, as a pilot project, we chose to target three sources, IRAS~17207-0014, IRAS~06035-7102, and IRAS~09022-361. These three sources maximised the detection of the water line within a short integration time.


 



\subsection{Observations}

We observed IRAS\,06035-7102 ($z=0.07946$), IRAS\,17208-0014 ($z=0.04281$), and IRAS\,09022-3615 ($z=0.05964$) with APEX SEPIA-660 Band 9 receiver (Project: E-0103.B-0471A-2018, PI: C. Yang) operating at the observing frequency range of $\sim$\,690--725\,GHz. The APEX beam size at this frequency is $9\farcs5$. Observations were carried out between May 11-12, June 03, 22-24, and August 21-22, 27-29 in 2019. The ON-OFF observations were performed in the wobbler-switching symmetric mode with an amplitude of 40\,arcsecs and a wobbling rate of 1.5\,Hz. The Doppler correction to account for the motion of the Earth was applied during the observations. R-Dor and L2-Pup were used for pointing and focus calibration. The focus was checked at the beginning of each observing session, and pointing was checked every approximately 1.5\,hours. Spectral setup covers 8\,GHz per sideband and a separation of 8\,GHz between the upper- and lower-side bands. The observation was centred on the redshifted line frequency of interest, and we requested a line peak-to-rms ratio of S/N$>$5 in velocity channels of $\sim$50\,km\,s$^{-1}$. The precipitable water vapour (PWV) during the observations varied between 0.4 and 0.9\,mm, corresponding to typical atmospheric transmission of 60\% and 20\%, respectively. The total integration times ranged between 4 and 15\,hrs per target (Table~\ref{tab:lineparameters}).
 
\subsection{Data reduction} 

The data were reduced using the Grenoble Image and Line Data Analysis Software 
(GILDAS)\footnote[1]{\url{https://www.iram.fr/IRAMFR/GILDAS/}} - Continuum and Line Analysis Single-dish Software 
(CLASS) package. For each target, we collected all the scans for different observing dates. For each observing date, 
we checked all the individual scans and discarded those affected by anomalous noise levels or weather conditions as 
noted by the log-file of the observations. 
For each scan, we masked a velocity window from $-$500 to $+$500\,km\,s$^{-1}$ centred at the expected line 
observing frequency to fit a second-order polynomial subtraction, which accounts better for the baseline shapes, to remove the baseline out of the spectra. 
Then we combined all of the individual spectra, using noise-based weighing and aligning them at the frequency of interest. 
We have also explored the baseline subtraction using first-order polynomials and did not find any major differences compared with the second-order polynomial fits.


\begin{figure}
	\includegraphics[width=\columnwidth]{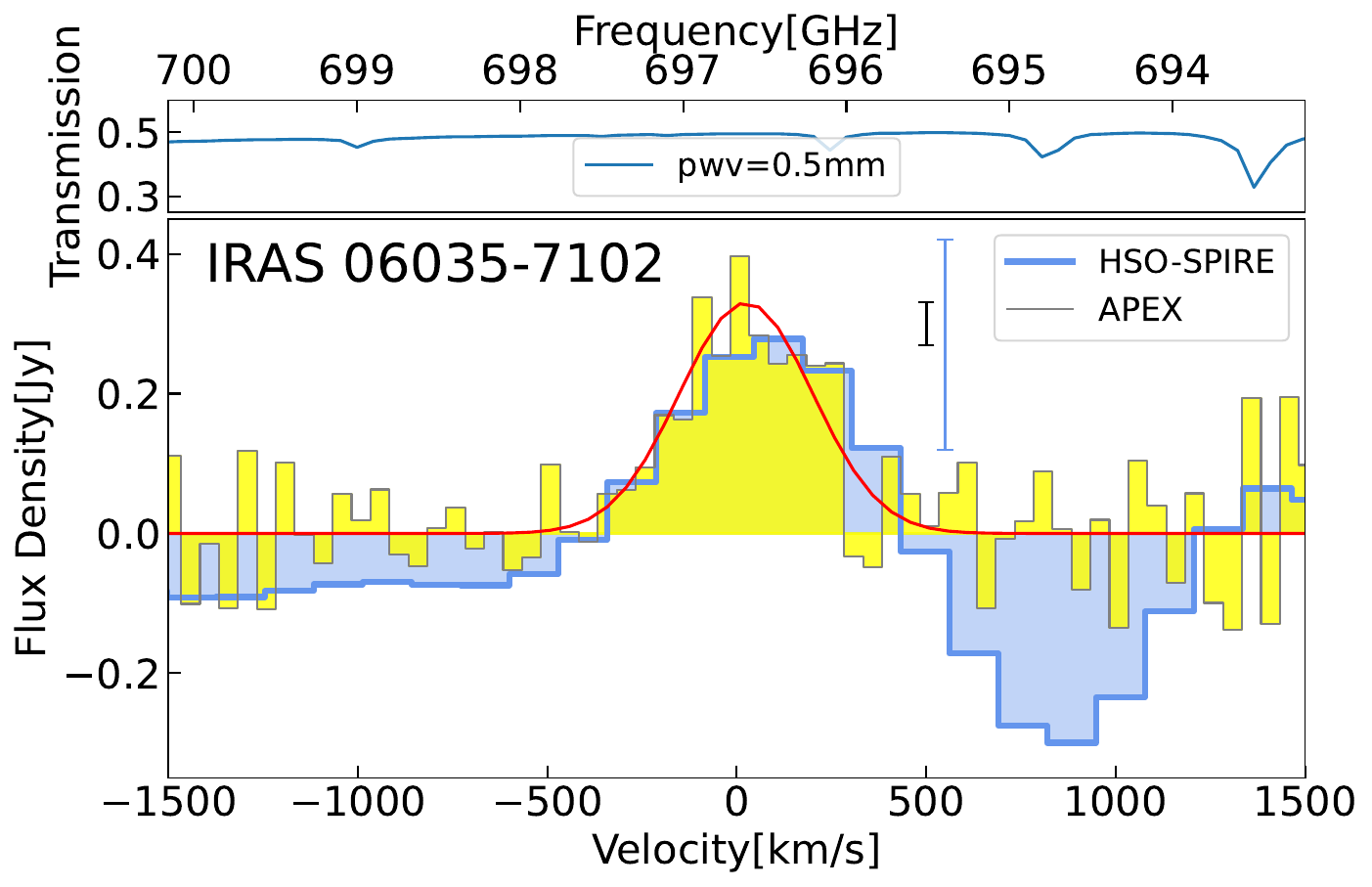}
	\includegraphics[width=\columnwidth]{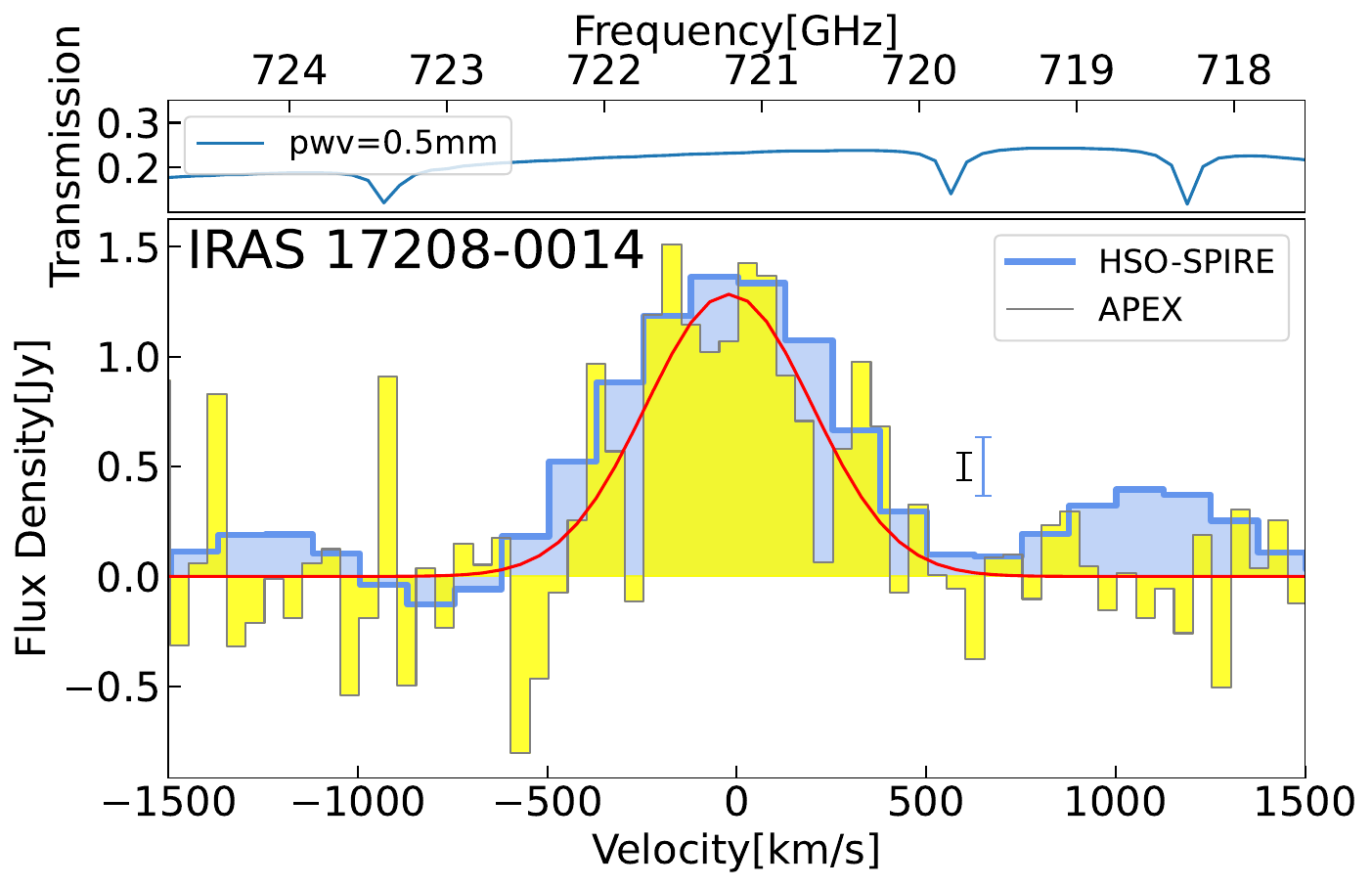}
	\includegraphics[width=\columnwidth]{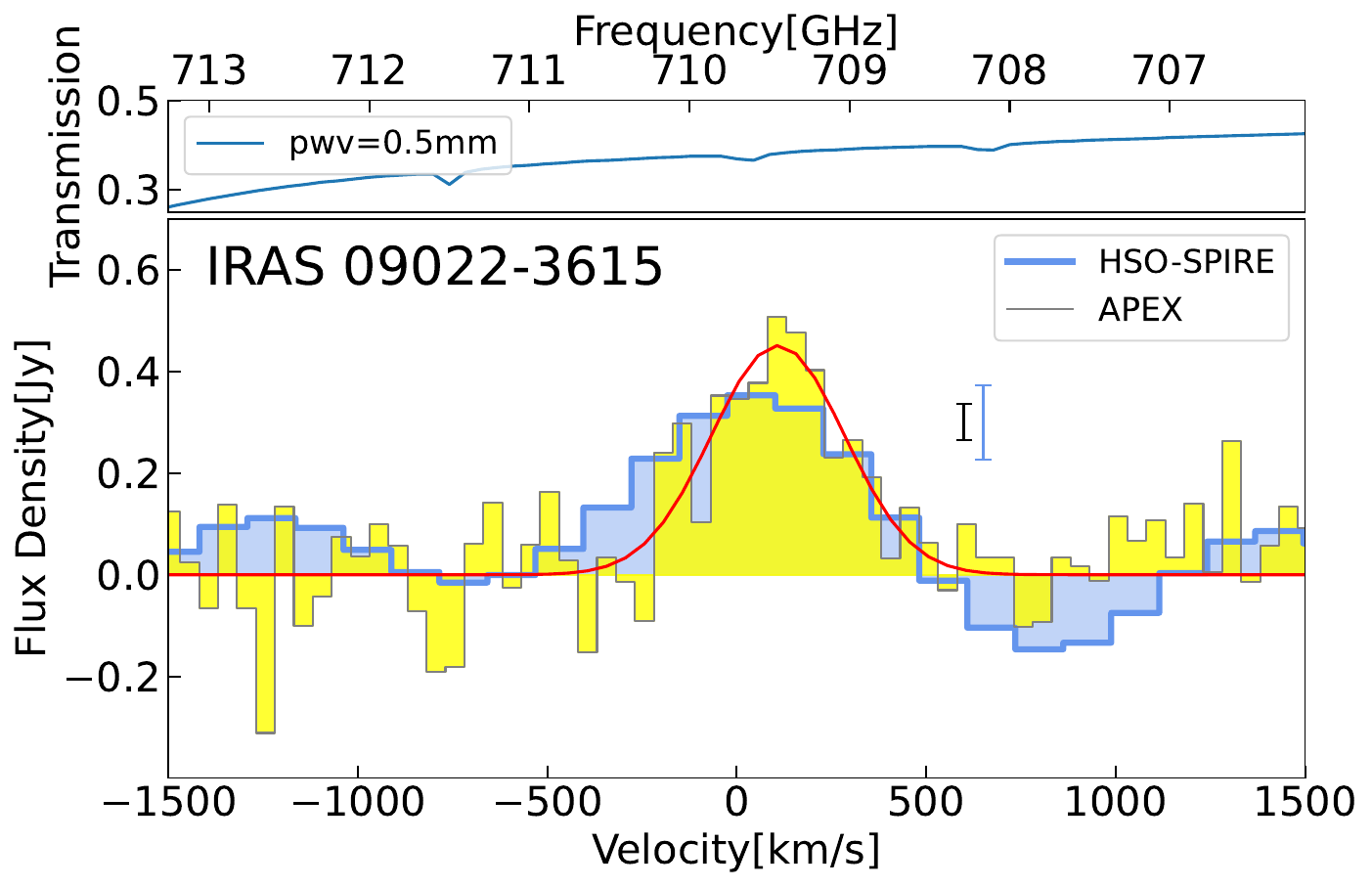}
    \caption{Redshifted para-H$_2$O(2$_{11}$–2$_{02}$) at 752.033\,GHz emission line for the three ULIRGs
    presented in this work. The spectra have been centred according to the redshift in Table~\ref{tab:lineparameters}.  {\color{black}The histograms show the binned} APEX spectra, while the red lines correspond to
    Gaussian fits.  
    Overplotted are the unapodized {\it Herschel} SPIRE-FTS spectra in blue. Error bars show a comparison 
    between uncertainties within a bandwidth of 40 GHz, in both spectra after considering a convolution of the APEX spectra with the SPIRE-FTS instrumental line spread function. The upper panel shows the atmospheric transmission for SEPIA 660 at PWV of 0.5 mm at an elevation of 55$^\circ$.}
    \label{fig:spectra}
\end{figure}

To display the spectra, we use a spectral resolution of 50\,km\,s$^{-1}$ channel width (see Figure~\ref{fig:spectra}). 
Considering the antenna temperature $T^*_\mathrm{A}$ corrected for the 
atmospheric attenuation, the forward efficiency and signal band gain, we derive flux densities
using a conversion factor of 75\,$\pm$\,6\,Jy\,K$^{-1}$ measured during the observations 
period with SEPIA-660\footnote[2]{\url{https://www.apex-telescope.org/telescope/efficiency/}}. 
The measured noise root mean squared (RMS), for all three sources,
ranges between 77 and 340 mJy\,beam$^{-1}$. 
The final spectra were exported to Python for the remaining analysis. 
IRAS\,17208-0014 became the brightest source, so less integration 
time was needed to reach the requested signal-to-noise ratios (see Table~\ref{tab:lineparameters}).

\subsection{SED photometry}  
\label{sed_section}
Previous analyses indicate that the p-\htot211202\ is likely to be dominated by far-IR pumping excitation by the photons at 101\,$\mu$m \citep[e.g.][]{Gonzalez2022}. If this is the case for our sources, we would expect a tight correlation between \hto\ emission and far-IR dust continuum.
{\color{black} 
In order to further explore this, we re-analyzed the 
{\it Herschel} Photodetector} Array Camera and Spectrometer (PACS) 100\,$\mu$m emission of IRAS\,17208-0014
(Obslist:1342241375, 1342241376) and IRAS\,09022-3615 (Obslist:1342233593, 1342233594). These were retrieved from the {\it Herschel} Public Archive (program OT1, PI: D. Sanders). Each galaxy was observed in cross-scanning mode at 20\,arcsec\,s$^{-1}$ and {\sc unimap} projected maps were considered. Since PACS simultaneously observes at 100 $\mu$m and at 160 $\mu$m, we performed aperture photometry at both wavelengths following the pointSourceAperturePhotometry script using the {\it Herschel} Interactive Processing Environment (HIPE) version 15.0.1 software \citep[][]{2010Ott}.  
We measured the flux at different aperture radii {\color{black} (up to $\sim$\,4\,$\times$ full width at half maximum, FWHM)} and checked the aperture at which the encircled energy fraction stabilizes. This profile is compared to the point spread function profile finding that both behave similarly, concluding that the two far-IR sources are point-like at the {\it Herschel} PACS beam (FWHM 
$\sim$\,6$\farcs$7\,$\times$\,6$\farcs$9 for 100\,$\mu$m and 
$\sim$\,10$\farcs$6\,$\times$\,12$\farcs$1 for 160\,$\mu$m). 
The photometric error is estimated using six equal apertures placed in the background around the source. The results for this aperture photometry are in agreement with those presented by \citet{Chu2017}. 

Using photometric points from the Wide-field Infrared Survey Explorer (WISE) 22\,$\mu$m, the Infrared Astronomical Satellite (IRAS) 60\,$\mu$m, PACS 70\,$\mu$m and {\it Herschel} SPIRE 250, 350, and 500\,$\mu$m, for all three sources we construct the Spectral Energy Distribution (SED) to estimate their total infrared luminosities \ir (see 
Table~\ref{tab:photometry} for the details).
 


\subsection{Ancillary submm data}

Public ALMA observations of CO($J$\,=\,4--3) and 630\,$\mu$m dust
continuum were retrieved for IRAS\,09022-3615 and 
IRAS\,17208-0014 (project ID: 2018.1.00994.S, PI: T. Michiyama). For IRAS 06035-7102, we retrieved data from HCN($J$\,=\,2--1) (project ID:2017.1.00022.S, PI: M. Imanishi) and 
from CO($J$\,=\,3--2) and 850\,$\mu$m dust
continuum (project ID:2018.1.00503.S, PI: A. Gowardhan). 
All observations were reduced using the standard pipeline 
calibration scripts provided by ALMA with the Common Astronomy Software Applications 
\citep[CASA;][]{TheCASATeam_2022} for the respective cycles. After obtaining the calibrated measurement sets, the 
imaging process was done with the task TCLEAN. We created an initial 
datacube where we identified the line-free channels. For the continuum subtraction, we used the task 
UVCONTSUB. We ran the cleaning process down to 1\,$\sigma$ using interactive masks located at the
source positions. The spectral resolution of the cubes is set to 50\,km\,s$^{-1}$, using a natural weighting and a
primary-beam corrected map. To better compare the p-\htot211202 line to the ALMA 
CO($J=4\text{--}3$), CO($J=3\text{--}2$), and HCN($J=2\text{--}1$) observations,
the ALMA data cubes were degraded in spatial resolution down to the same APEX beam 
using the task IMSMOOTH. The final ALMA spectra used for comparison are extracted at the convolved central 
brightest pixel in Jy\,beam$^{-1}$ units (Figure~\ref{fig:spectra}). 

\begin{figure*}
	\includegraphics[scale=0.28]{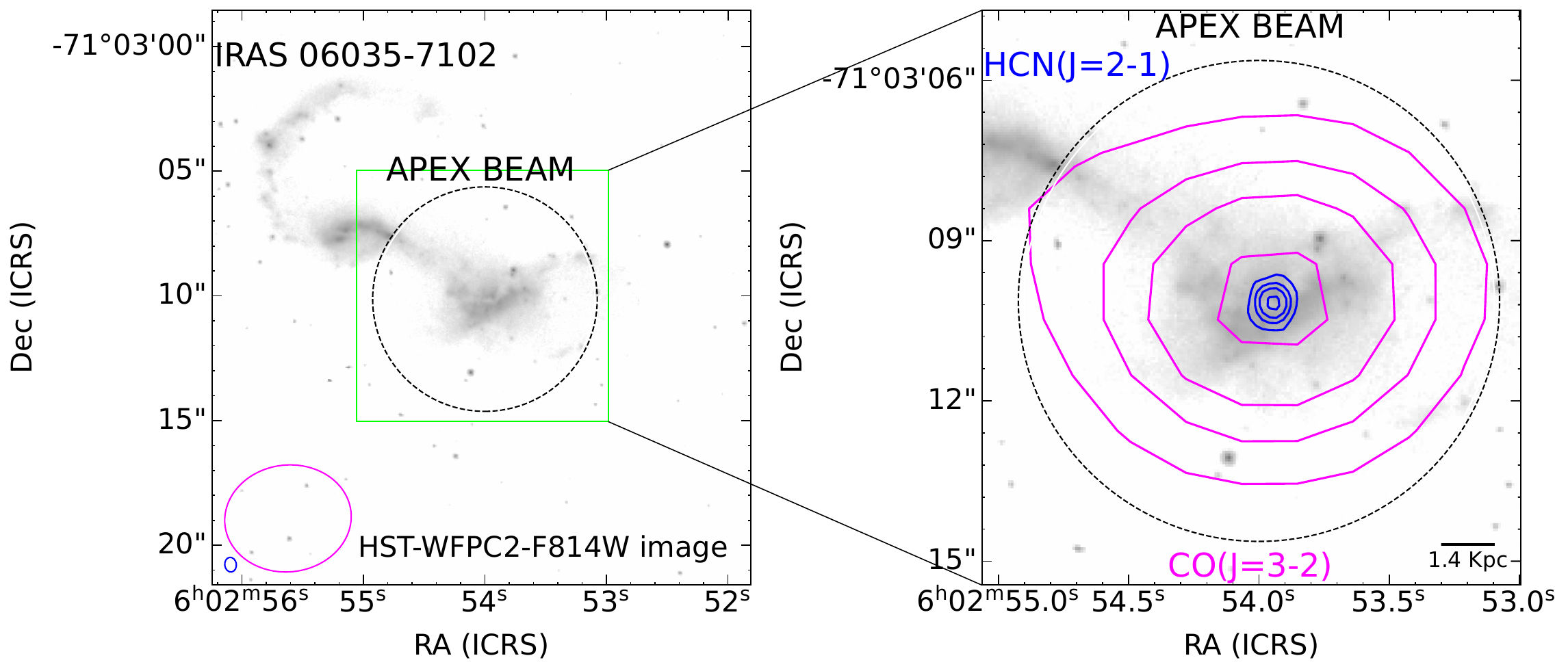}
	\includegraphics[scale=0.30]{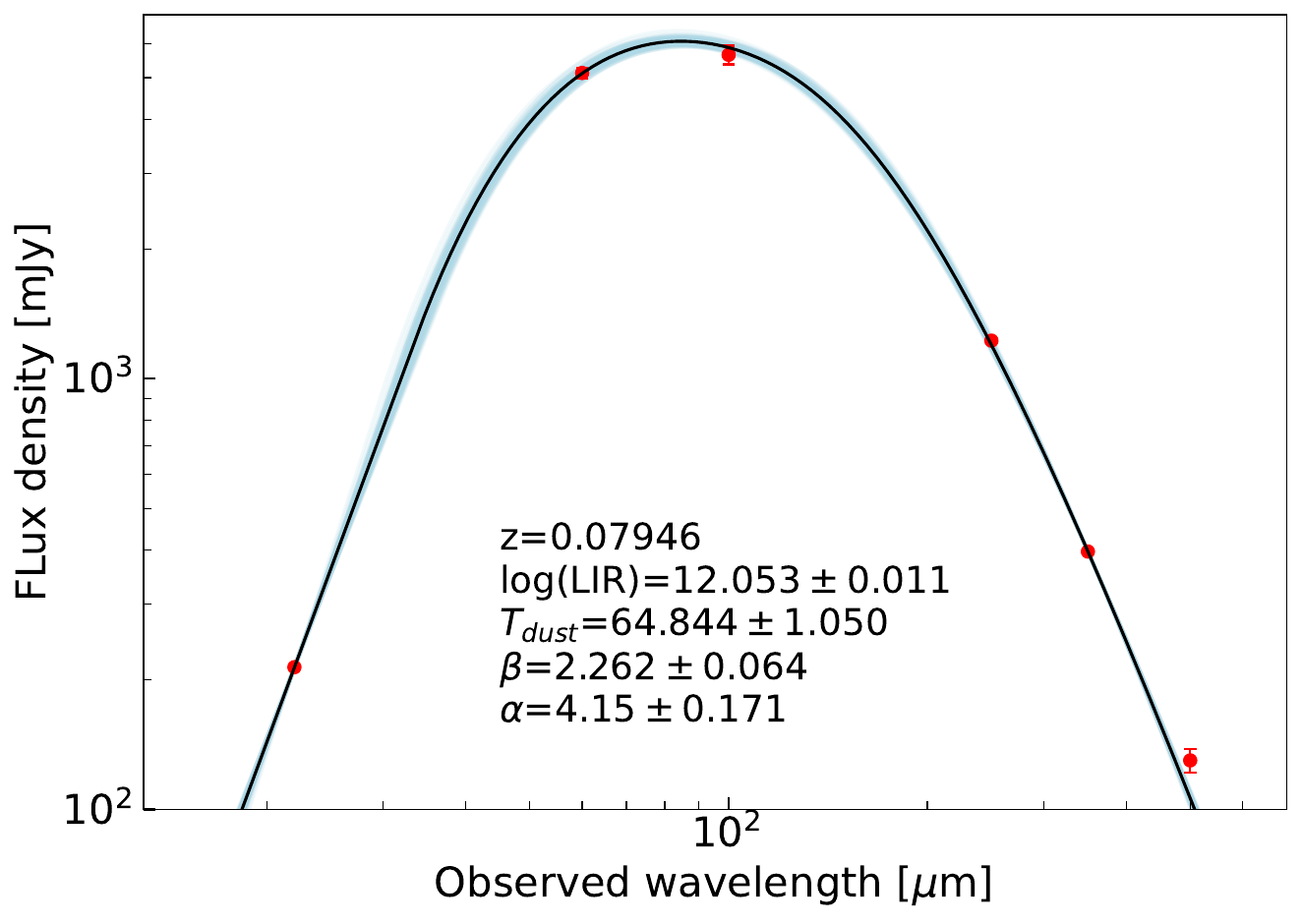}
	\includegraphics[scale=0.28]{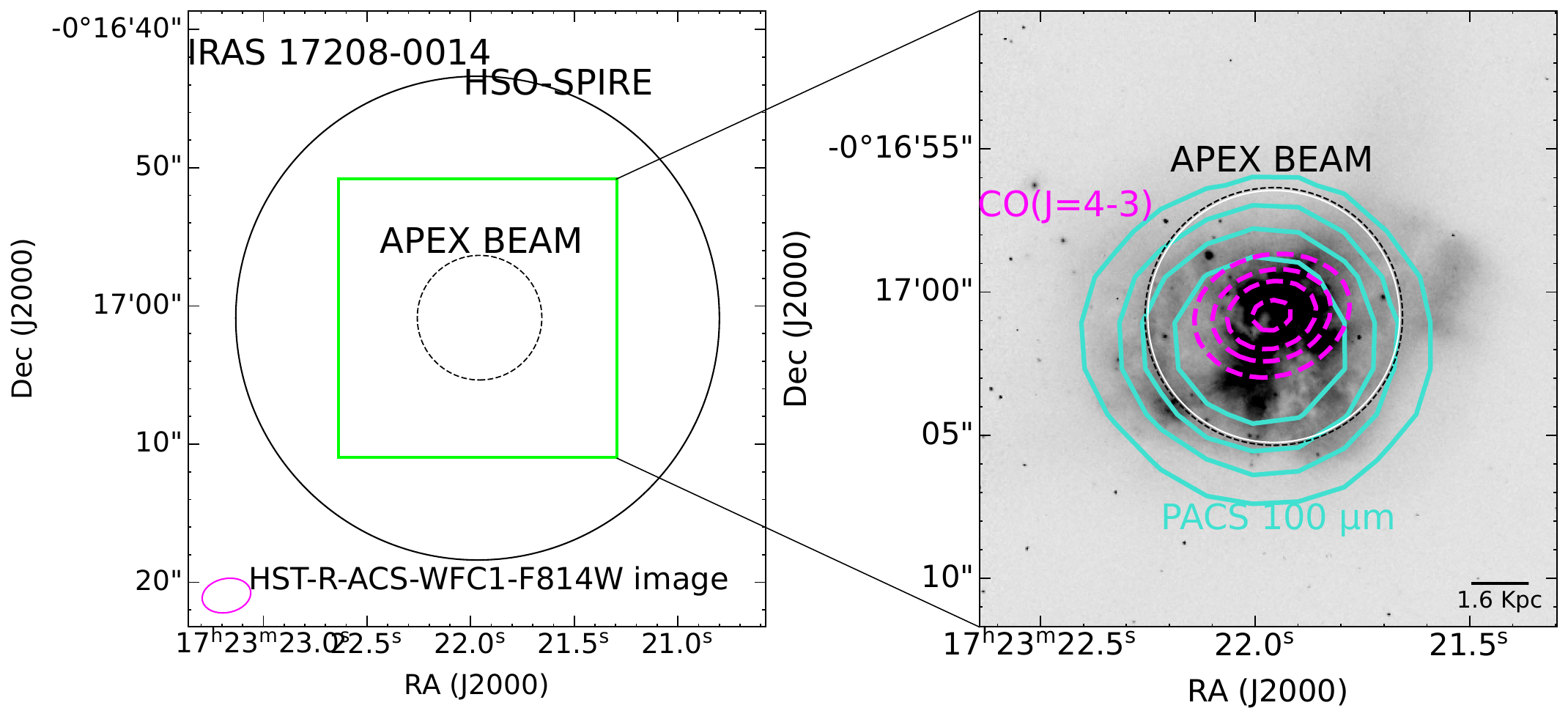}
	\includegraphics[scale=0.30]{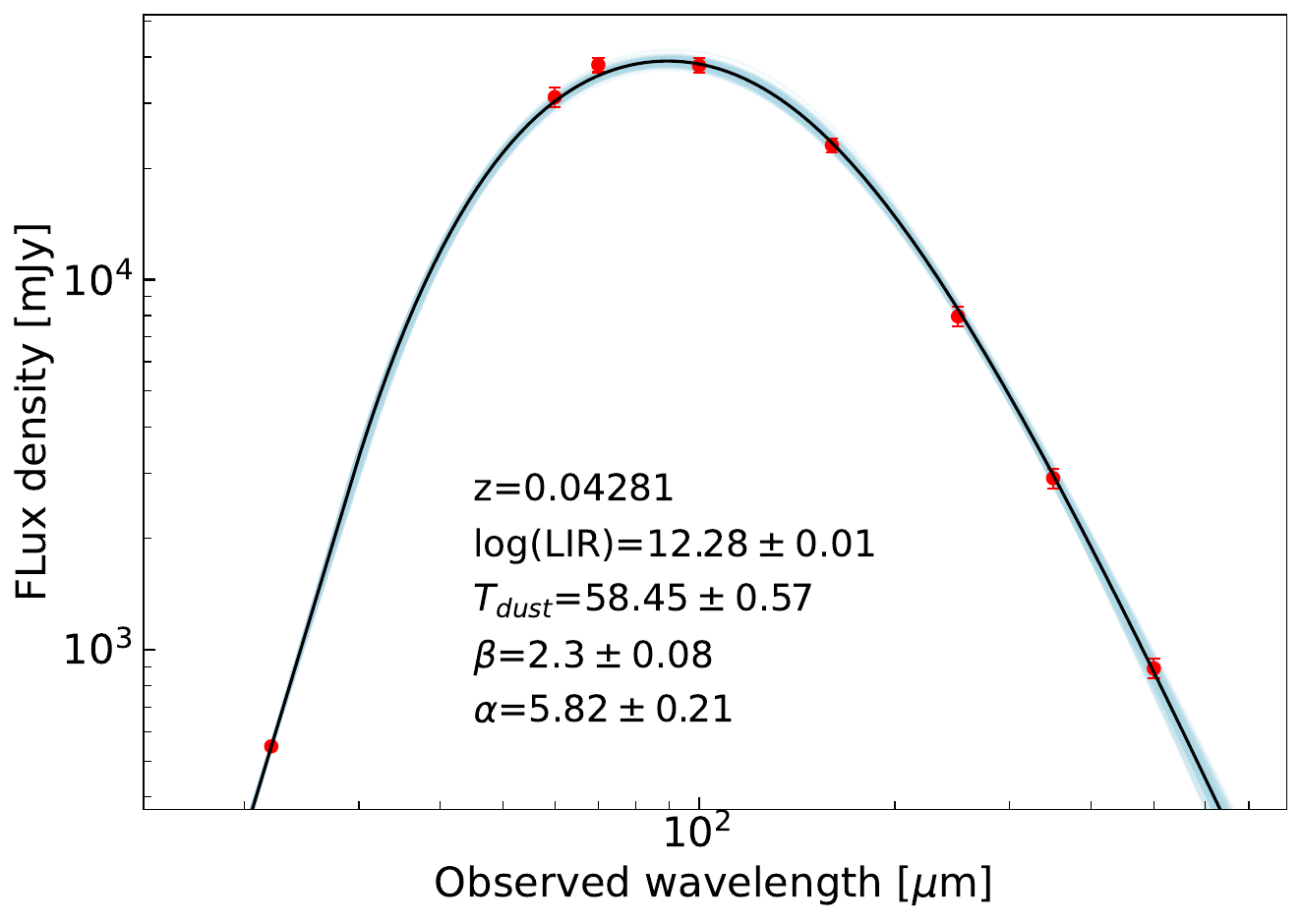}
	\includegraphics[scale=0.28]{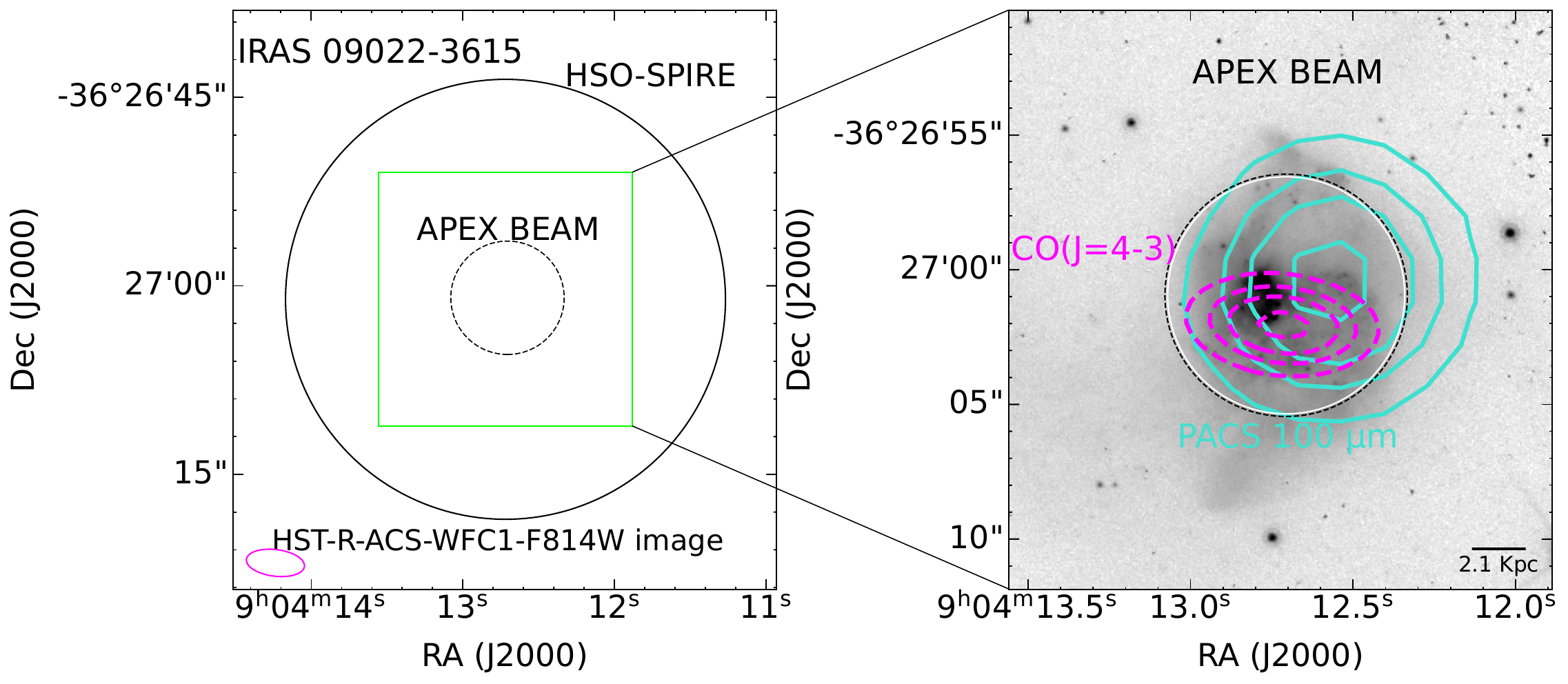}
	\includegraphics[scale=0.30]{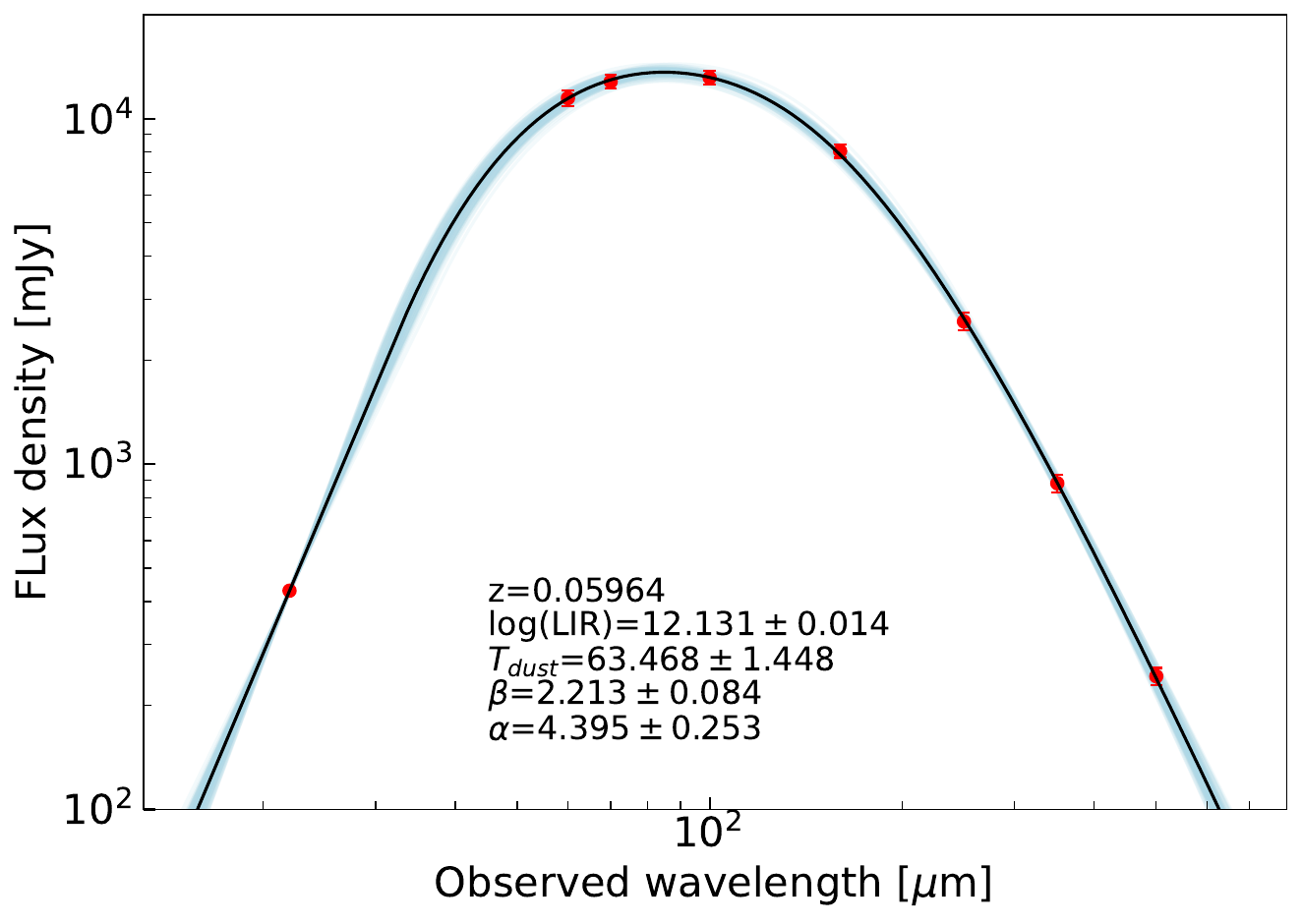}
    \caption{{\it Left}: Hubble Space Telescope R-band images of the three ULIRGs presented in this work. We use a logarithmic stretch function to enhance the faint emission. The beams of {\it Herschel} SPIRE (solid circle) and APEX (dashed line) are overlaid. {\color{black}{\it Center}: The contours represent the 20$\%$, 40$\%$, 60$\%$, 90 $\%$ of the maximum peak intensity at 100 $\mu$m from {\it Herschel} PACS (cyan) and CO($J$\,=\,3-2) (magenta), CO($J$\,=\,4-3) (magenta) or HCN($J$\,=\,2-1) (blue) from ALMA}. These maps are at their native resolution and their beams are on the lower side of the left hand side figure.  {\it Right}: The observed far-IR SED for each galaxy, including photometry from WISE, IRAS, and {\it Herschel} (red points). The fitted SED is shown as a solid black line, while the blue shadow shows $1\sigma$ uncertainties.}
    \label{fig:BEAM}
\end{figure*}

\section{Results and Discussion}

\setlength{\tabcolsep}{0.75em}
\begin{table*}  
	\caption{Observed \htot211202 line parameters}
	\label{tab:lineparameters}
	\begin{tabular}{lcccccccccr} 
		\hline
		N &Source & RA (J2000) & Dec (J2000) & z & $\nu_\mathrm{obs}$ & $D_\mathrm{L}$ & FWHM$_\mathrm{H_2O}$ & $S_\mathrm{H_2O}\,\Delta$v & 
            $L_\mathrm{H_2O}$          & $t_\mathrm{int}$\\
		   &       &            &             &   & [GHz]              & [Mpc]          & [km\,s$^{-1}$]      & [Jy\,km\,s$^{-1}$]         & 
            [$\textup{L}_{\sun}/10^6$] & [h]             \\
		   &(1)&(2)&(3)&(4)&(5)&(6)&(7)&(8)&(9)&(10)     \\ \hline
		1&IRAS\,06035-7102  & 06:02:54.01 & -71:03:10.2 &0.07946 &696.7& 325.0&414.88$\pm$54.01 &145.73$\pm$16.32& 10.33$\pm$1.16 &12.3\\
		2&IRAS\,17208-0014 & 17:23:21.96 & -00:17:00.9 &0.04281&721.2&175.68 &515.90$\pm$68.61 &704.77$\pm$83.72& 15.64$\pm$1.86 &4.2 \\
	      3& IRAS\,09022-3615 & 09:04:12.71 & -36:27:01.0 &0.05964&709.7 &238.84&414.81$\pm$49.89 & 199.08$\pm$20.84&7.90$\pm$0.83 &14.3 \\
		\hline
	\end{tabular}
    \begin{tablenotes}[flushleft]
   \small
	 \item\textbf{Note:} (1) Source name. (2) Right ascension. (3) Declination. (4) redshift from the NASA/IPAC Extragalactic Database (NED). (5) Central observed frequency. (6) Luminosity distance in Mpc from \citet{Yang2013}. (7) The \htot211202\ FWHM. (8) The velocity-integrated flux density. (9) Line luminosity. (10) Total APEX integration time (including overheads). 
   \end{tablenotes}
\end{table*}

\subsection{Water line measurements}

\begin{table} 
	\caption{A comparison of the velocity integrated p-\htot211202 flux densities S$_\mathrm{H_2O}\Delta$v between our results and previous ones with {\it Herschel} SPIRE-FTS taken from the literature.}
	\label{tab:integrated_flux}
	\begin{tabular}{lccr} 
		\hline
		Source&APEX& HSO-SPIRE &HSO-SPIRE \\
			&[Jy km s$^{-1}$]&[Jy~km~s$^{-1}$]&[Jy~km~s$^{-1}$]\\
		&&\citet{Yang2013}&\citet{Pearson2016} \\%
	
		\hline
 1&145.73 $\pm$ 16.32&\(<411^*\)&\\ 
2&\( 704.77\pm 83.72\)&\( 697.98 \pm 88.70\)&\(654.23 \pm 57.40^{**}\)\\
 3&\( 199.08\pm 20.84\)&\( 184.65 \pm 52.32\)&\( 165.95 \pm 40.68^{**}\)\\
		\hline
\end{tabular}
\textbf{Note:}
$^*$ The value is a 3-$\sigma$ upper limit, although the original work reports a $<$\,3\,$\sigma$ tentative detection of $408.63\pm 136.98$~Jy~km~s$^{-1}$ with low significance. $^{**}$ The conversion factor from W~m$^{-2}$ to Jy~km~s$^{-1}$ is assumed $Q=3\times 10 ^{22}/\nu_\mathrm{rest}$[GHz].
\end{table}

Using APEX SEPIA Band-9 we detected the thermal transition of the water molecule 
p-\htot211202 in all three ULIRGs at $>8\sigma$ significance levels. 
For IRAS\,06035-7102, this is the first clear detection of p-\htot211202, 
superseding the 
tentative $\sim3\sigma$ detection reported by \citet{Yang2013}. 
The spectra of p-\htot211202 observed with APEX toward IRAS\,06035-7102, IRAS\,17208-0014, and IRAS\,09022-3615 are shown in Figure~\ref{fig:spectra}, {\color{black} in comparison with the low resolution {\it Herschel} SPIRE-FTS spectra (the full spectra can be seen in Figure~\ref{fig:SPIRE})}. {\color{black} We note that the zero velocity is defined by the redshift of the source, as indicated in Table \ref{tab:lineparameters}. We have identified a  velocity shift of 100 km/s in the case of IRAS\,09022-3615, which will be discussed in detail in Section 3.4.}  We confirm the line detection at a higher significance (at least a factor of two) and higher spectral resolution. The spectra show that p-\htot211202 is detected with a peak flux density of 
$0.45 \pm 0.09$\,Jy\,beam$^{-1}$ in IRAS\,09022-3615, 
$1.28 \pm 0.34 $\,Jy\,beam$^{-1}$ in IRAS\,17208-0014 and 
$0.33  \pm 0.08$\,Jy\,beam$^{-1}$ in IRAS\,06035-7102. The velocity-integrated flux densities and a comparison with previous values taken from the literature are presented in Table~\ref{tab:integrated_flux}. The values agree within the uncertainties, 
with the line fluxes and tentative limits measured with the {\it Herschel} SPIRE-FTS by \citet{Yang2013} and \citet{Pearson2016}. 

The instrumental line spread function of SPIRE-FTS is a function that redistributes the line flux along the frequency axis \citep[][]{2015Hopwood}, which correlates the channels across the line profiles. On the other hand, as SEPIA is a heterodyne instrument, it does not have this line spread function effect, offering independent channels and demonstrating its power for resolving the line profiles. Thus APEX not only provides an accurate measurement of the line flux but also provides a direct way to recover the true line profiles, revealing further information about the kinematics of the gas.

The high spectral resolution of APEX allows us to resolve the p-\htot211202 lines {\color{black}and determine the full width at half maximum (FWHM) in all three galaxies, which are 415, 516 and 415\,km\,s$^{-1}$ for} IRAS~06035-7102, IRAS~17208-0014 and IRAS~09022-3615, respectively (Table~\ref{tab:lineparameters}). Within the $9\farcs5$ APEX beam and a spectral resolution of 50\,km\,s$^{-1}$, the line profiles are consistent with a single Gaussian, facilitating clean measurements of their velocity-integrated flux densities. 

To compare the spectral noise between the APEX and the previous {\it Herschel} 
SPIRE-FTS measurements, we convolved the APEX spectra with the 
line spread function profile of SPIRE-FTS and then binned these spectra 
to match the spectral SPIRE-FTS resolution. The error bars (measured within 
a 40\,GHz bandwidth for both) obtained for the convolved/binned APEX and SPIRE-FTS are 
shown in Figure~\ref{fig:spectra}, 
demonstrating the higher significance reached by APEX to characterise 
the p-\htot211202 line emission.

To compute the H$_2$O luminosities, we follow the equation in \citet{Solomon2005}: 
\begin{equation}
    L_\mathrm{H_2O}=1.04\times 10^{-3} S_\mathrm{H_2O}\Delta v \nu_\mathrm{rest}(1+z)^{-1} D_\mathrm{L}^2 [L_{\sun}],
    \label{eq:luminosity}
\end{equation}

\noindent
where $S_\mathrm{H_2O}\Delta v$ is the velocity integrated flux density in units of 
Jy\,km\,s$^{-1}$, the rest-frame frequency of the line $\nu_\mathrm{rest}$ is related to the observed frequency $\nu_\mathrm{obs}$ as $\nu_\mathrm{rest}=\nu_\mathrm{obs}(1+z)$ is in GHz, $D_{\rm L}$ is the luminosity distance in Mpc and $z$ is the redshift. Derived luminosities are presented in Table~\ref{tab:lineparameters}.

In Figure~\ref{fig:BEAM}, we present a comparison between the APEX ($9\farcs5$) and {\it Herschel} SPIRE-FTS (35\arcsec) beam sizes. 
As both spectra show similar intensities, we suggest that 
most of the water emission comes from a compact region smaller than
the APEX beam. To look at the optical nature of the sources, we present images from the {\it Hubble Space Telescope} (HST) with the Advanced Camera for Surveys (ACS), the Wield Field Camera (WFC) and the Wide Field Planetary Camera (WFPC2) from programs 6346 (WFPC2, PI: K. Borne) and 10592 (ACS, PI: A. Evans, see \citealt{Kim_2013}) in the $I$ filter (F814W, $\lambda = 8333$\r{A}). The $I$-band images of our galaxies have a large field of view ($202\arcsec \times 202''$) and capture the detailed structure of the galaxies and the full extent of each ULIRG interaction.  
We can see that IRAS\,06035-7102 is an extended interacting double system, separated by a distance of $\sim$\,10$''$ (in WFC2 image; \citealt[][]{2008Arribas}). 
IRAS\,17208-0014 is an advanced merger with two tidal tails toward the southeast and northwest.
IRAS\,09022-3615 is a late-type merger presenting a tidal tail. 
While in IRAS\,06035-7102 and IRAS\,17208-0014, the ALMA mid-$J$ CO and HCN(2--1) maps show that the molecular gas is concentrated towards the nuclear region where the far-IR emission peaks, the CO(3--2) emission in IRAS\,09022-3615 locates offset from the peak of the HST image and the far-IR dust emission, indicating complex ISM structure and kinematics. Our APEX beams are well aligned with the bulk of the compact molecular gas and dust emissions, which should trace the total \hto\ emission from our targets. 

 \begin{figure}
	\includegraphics[width=\columnwidth]{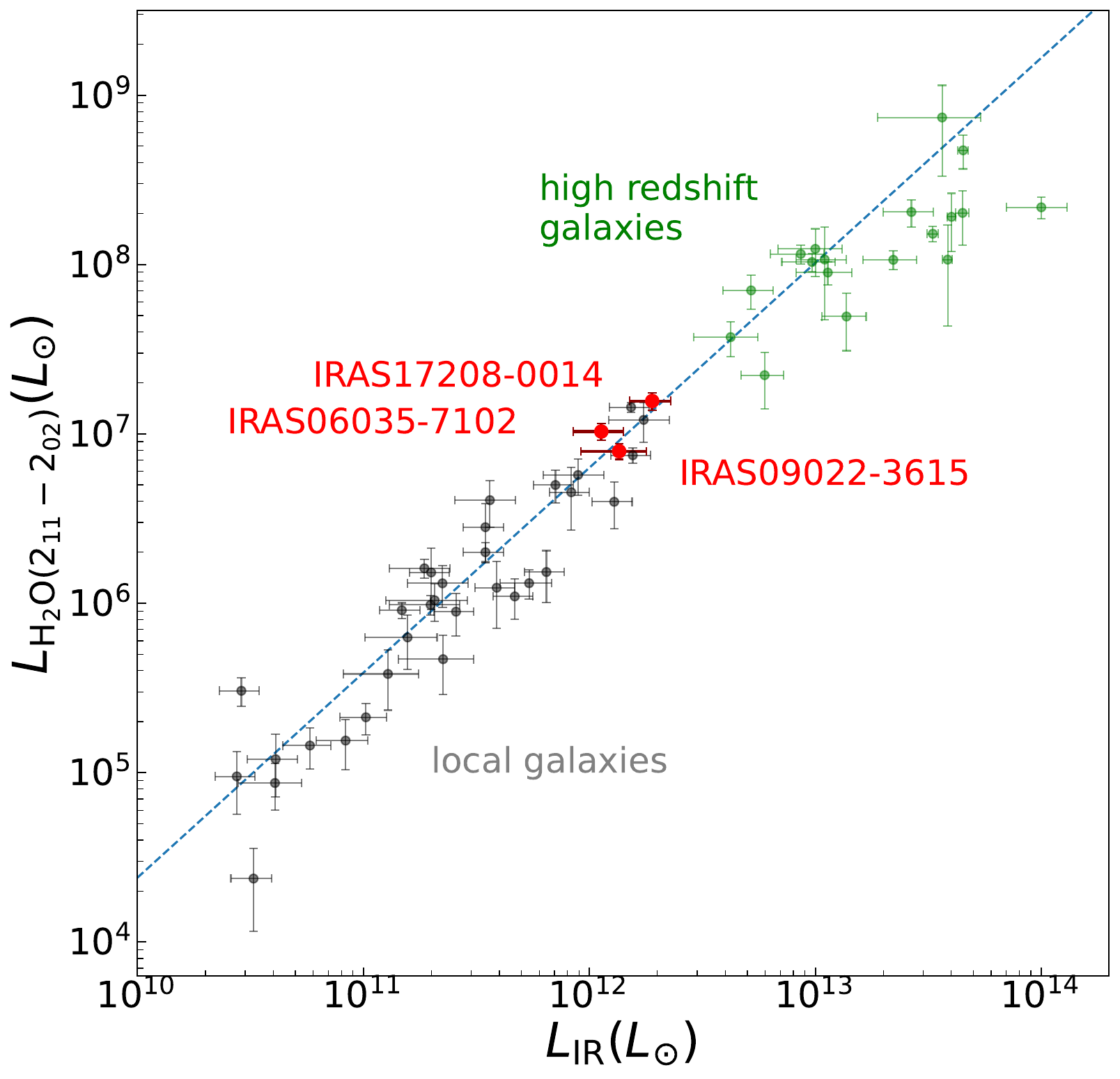}
    \caption{Correlation between $L_{\rm H_2O}$ and \ir in local (grey) and high redshift (green and blue) galaxies. The  linear correlation by \citet[][]{Yang2013}{}{} is shown as a blue dashed line at $L_\mathrm{H_2O}/$\ir$=$ 5.53 $\times$ 10$^{-6}$. {\color{black} Shown in red are the derived values for the three galaxies studied in this paper. The local galaxy sample and the high-$z$  ULIRGs are taken from \citet[][]{Yang2013, Yang2016,Berta2023}{}{}}}
    \label{fig:lh2o_lir}
\end{figure}


\subsection{{\color{black}The IR luminosities}}

To estimate the total infrared luminosity (\ir), we model the SED of galaxies between 8 and 1000\,$\mu$m. 
We use the Monte Carlo InfraRed SED (MCIRSED; \citealt{Drew2022}), which fits the data with a function consisting of a mid-infrared (mid-IR) power law and a far-IR modified blackbody. The model has the form 
\begin{equation}\label{eq:1}
  S(\lambda) =
      \begin{cases}
        N_\mathrm{pl}\lambda^{\alpha} & \text{: } \frac{\partial\log S}{\partial\log \lambda} > \alpha \\
        
        \frac{N_\mathrm{bb} \left(1-e^{-(\lambda_{0}/\lambda)^{\beta}}\right) \lambda^{-3}}{e^{hc/\lambda kT}-1} & \text{: } \frac{\partial\log S}{\partial\log \lambda} \leq \alpha,
      \end{cases}
\end{equation}
where $\lambda$ is the rest-frame wavelength, $\alpha$ is the mid-IR power-law slope, $N_\mathrm{pl}$ and $N_\mathrm{bb}$ are normalization constants, $\lambda_0$ is the wavelength where the dust opacity equals unity, $\beta$ is the dust emissivity index, $T$ is the luminosity-weighted characteristic dust temperature, $h$, $k$ and $c$ are Planck, Boltzmann and the speed of light constants. The two functions connect at the wavelength where the slope of the modified blackbody is equal to the slope of the power law function. In our model, $\alpha$ characterises the hot dust emission and is constrained to be between 0 <$\alpha$< 6, while on the other hand, $\beta$ controls the slope in the Rayleigh-Jeans regime and it is constrained to be between 0.5 and 5.0. 
In Figure~\ref {fig:BEAM}, we show the SED fitting and the estimated fitted values. 
The value of $\alpha$ for the three galaxies is steep ($>4.0$), consistent with a lack of significant hot dust emission
coming from a buried powerful AGN. The value of the emissivity index, less than 2.5, is in agreement with most values in local star-forming galaxies. As expected, the infrared luminosities are higher than $10^{12}$\,$L_\odot$, consistent with being ULIRGs. 


\subsection{The $L_{\rm IR}-L_{\rm H_2O}$ correlation}

In Figure~\ref{fig:lh2o_lir}, we present the correlation between $L_\mathrm{H_2O}$ and \ir as found in local and high redshift galaxies. Our sample is highlighted in red and is part of the brightest low redshift ULIRGs. Previous studies have shown that the \ir-$L_\mathrm{H_2O}$ correlation is slightly superlinear \citep[e.g.][]{Omont2013, Yang2013, Jarugula2019} compared to other water lines with the exception of H$_2$O(2$_{11}$--2$_{02}$), which is also excited by 101\,$\mu$m photons. This correlation opens the possibility of using water as a tracer of the far-IR emission and suggests that the IR pumping mechanism is the main responsible for the excitation of water molecules producing the p-\htot211202 line at 752GHz. Nevertheless, studies of local galaxies show that the low-J ($J<3$) H$_2$O lines {\color{black}can also be both collisionally and far-IR excited} \citep[][]{Liu2017}. One simple approach to hint light on the nature of the exciting mechanism of the p-\htot211202 is to compare the spatial distribution of the water emission to that of the dust continuum near $101\,\mu$m, compared to other molecular gas tracers, which are well known to be collisionally excited, such as CO($J$\,=\,3--2), CO($J$\,=\,4--3) or HCN($J$\,=\,2--1). However, given that our APEX observations are only {\color{black} consistent with single pointings, we here compare the line profiles as an indication of a common nature, assuming that kinematic structures are associated with similar spatial distributions.}


\begin{figure*}
    \includegraphics[scale=0.35]{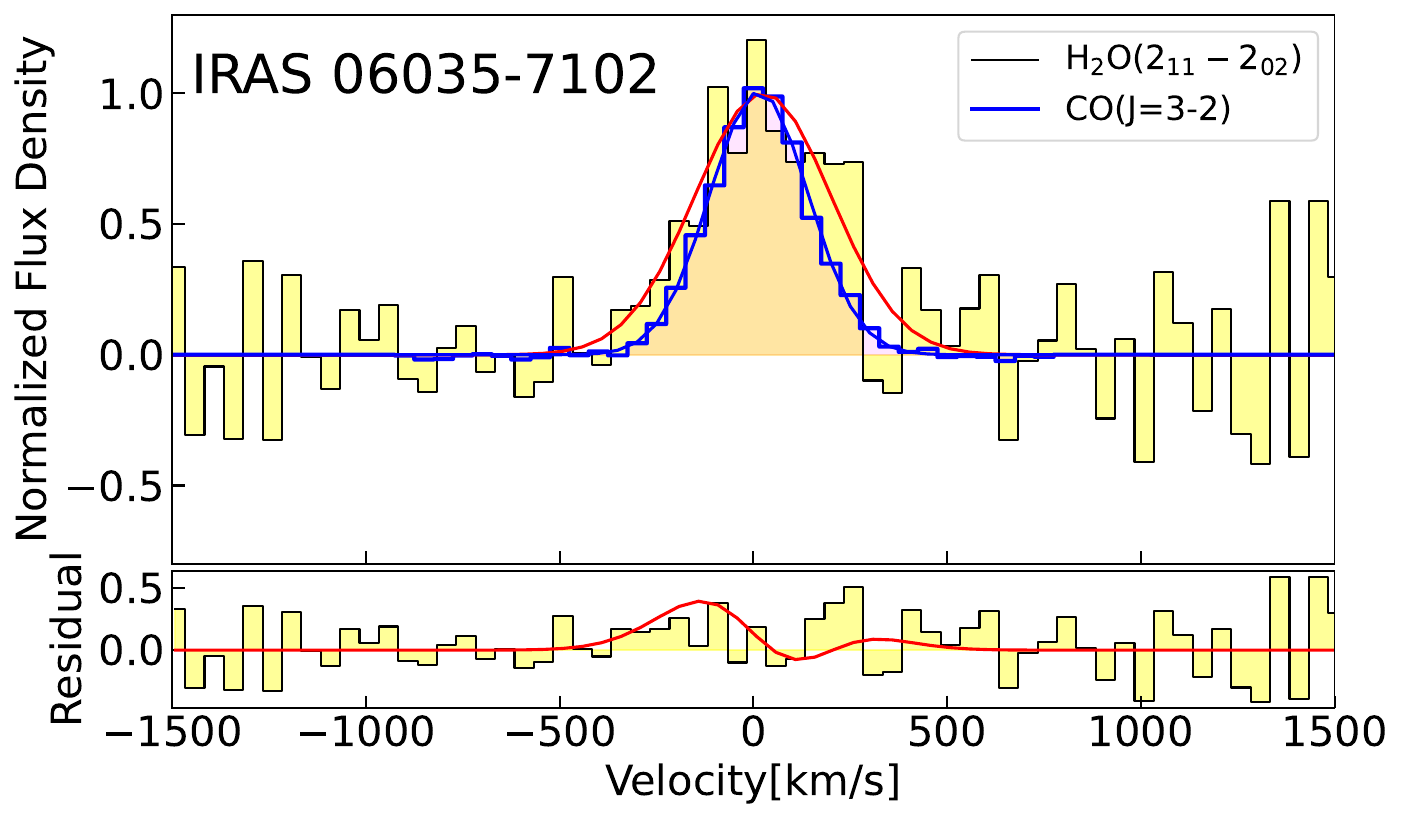}
	\includegraphics[scale=0.32]{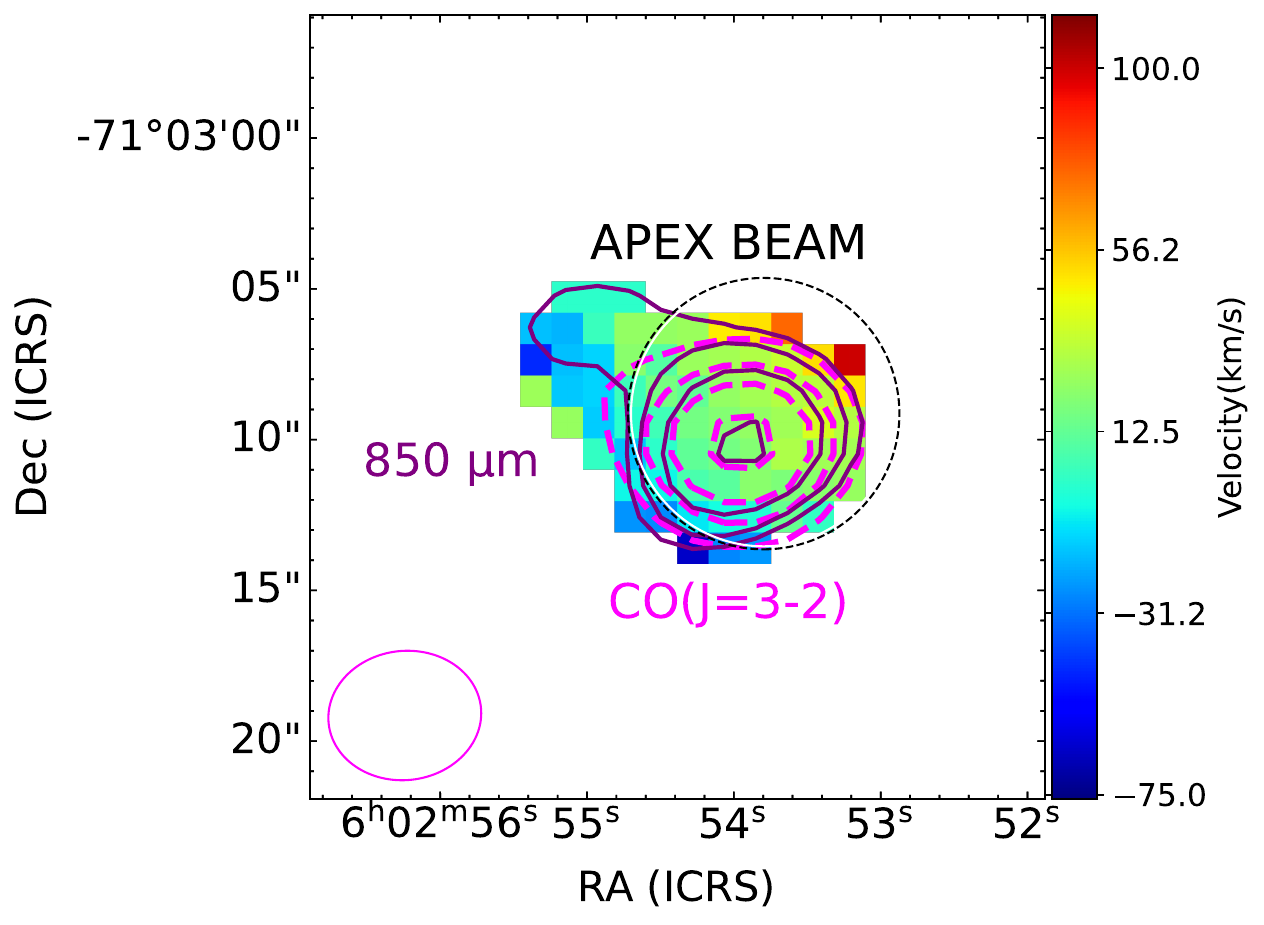} 
    \includegraphics[scale=0.35]{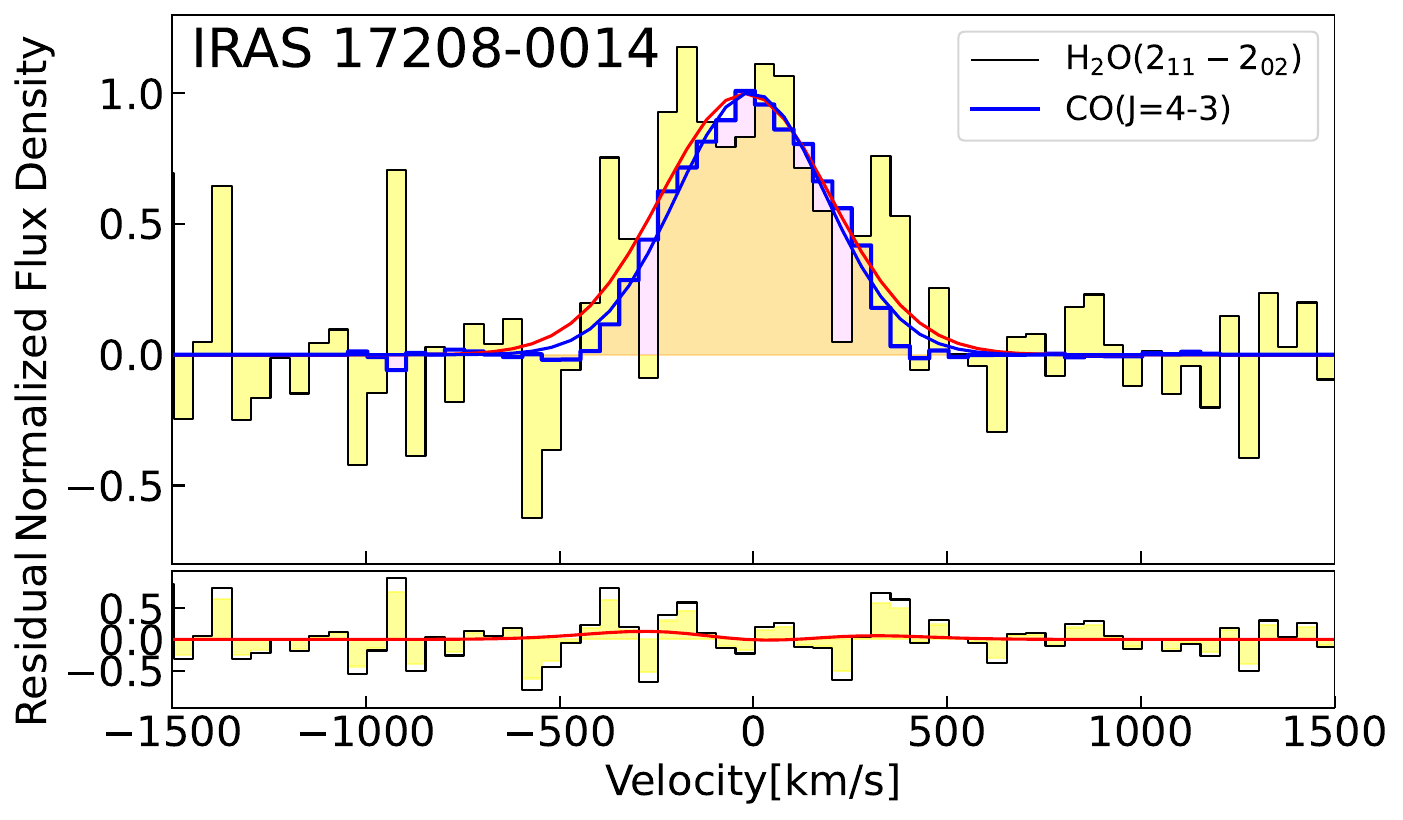}
	\includegraphics[scale=0.32]{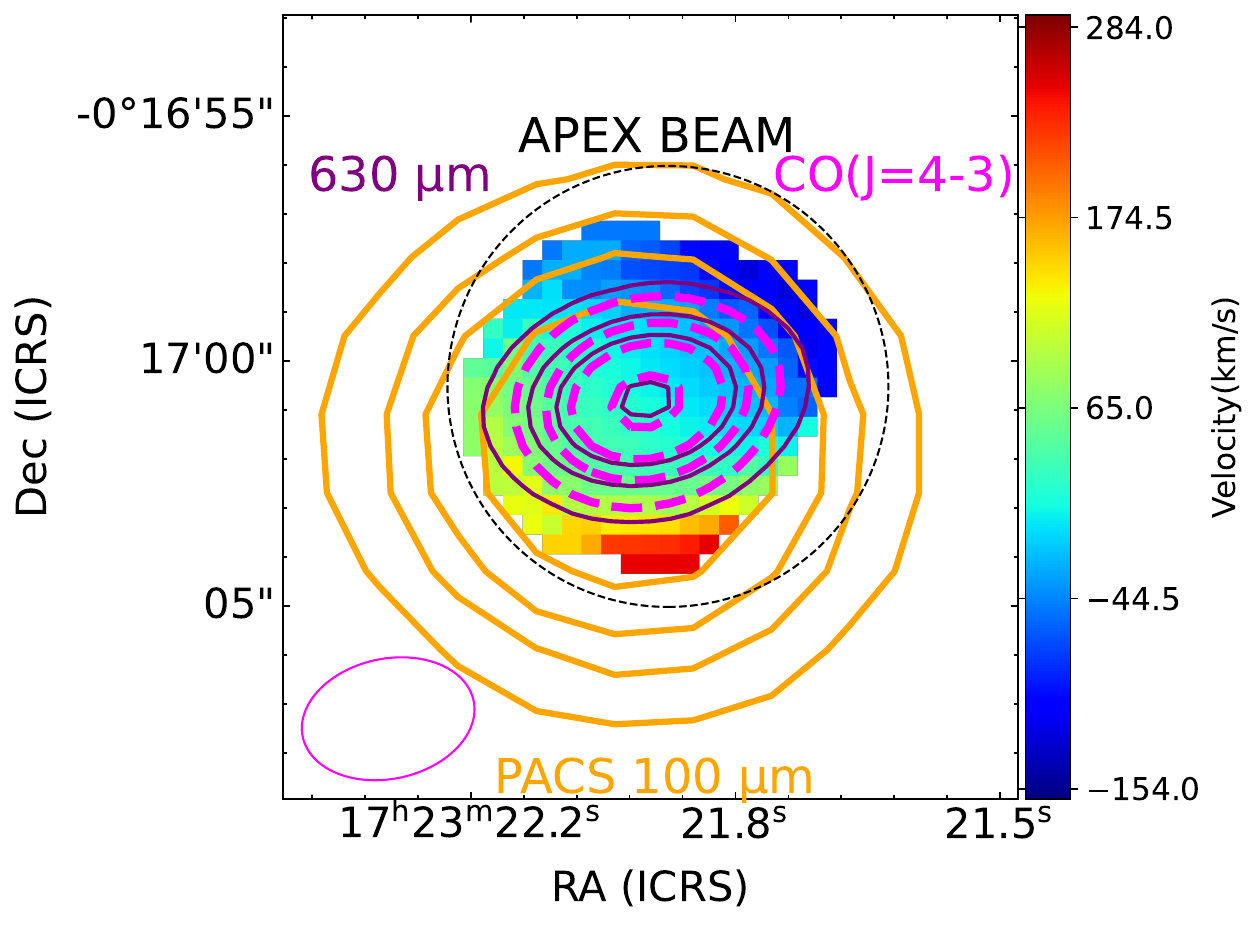} 
     \includegraphics[scale=0.35]{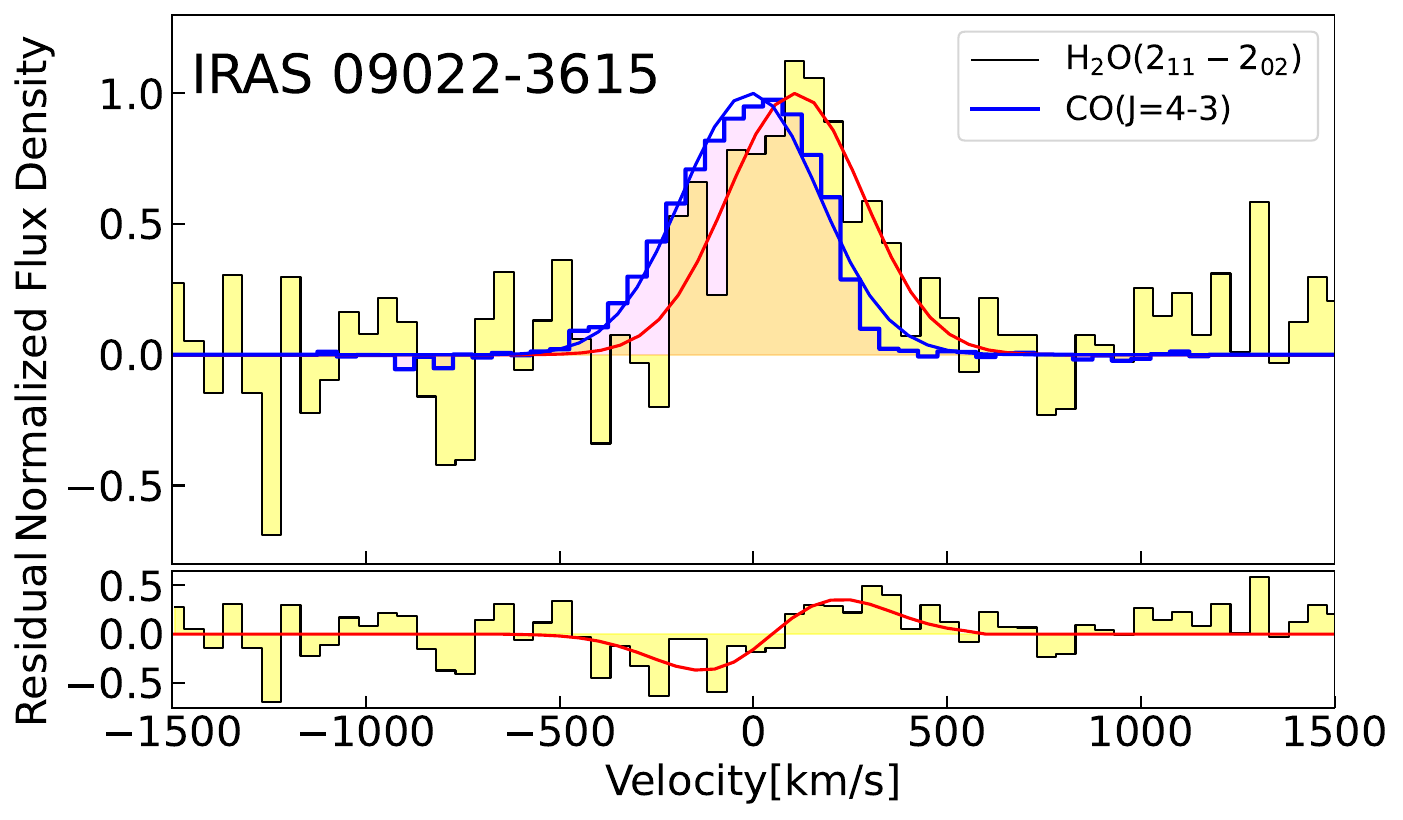}
	\includegraphics[scale=0.33]{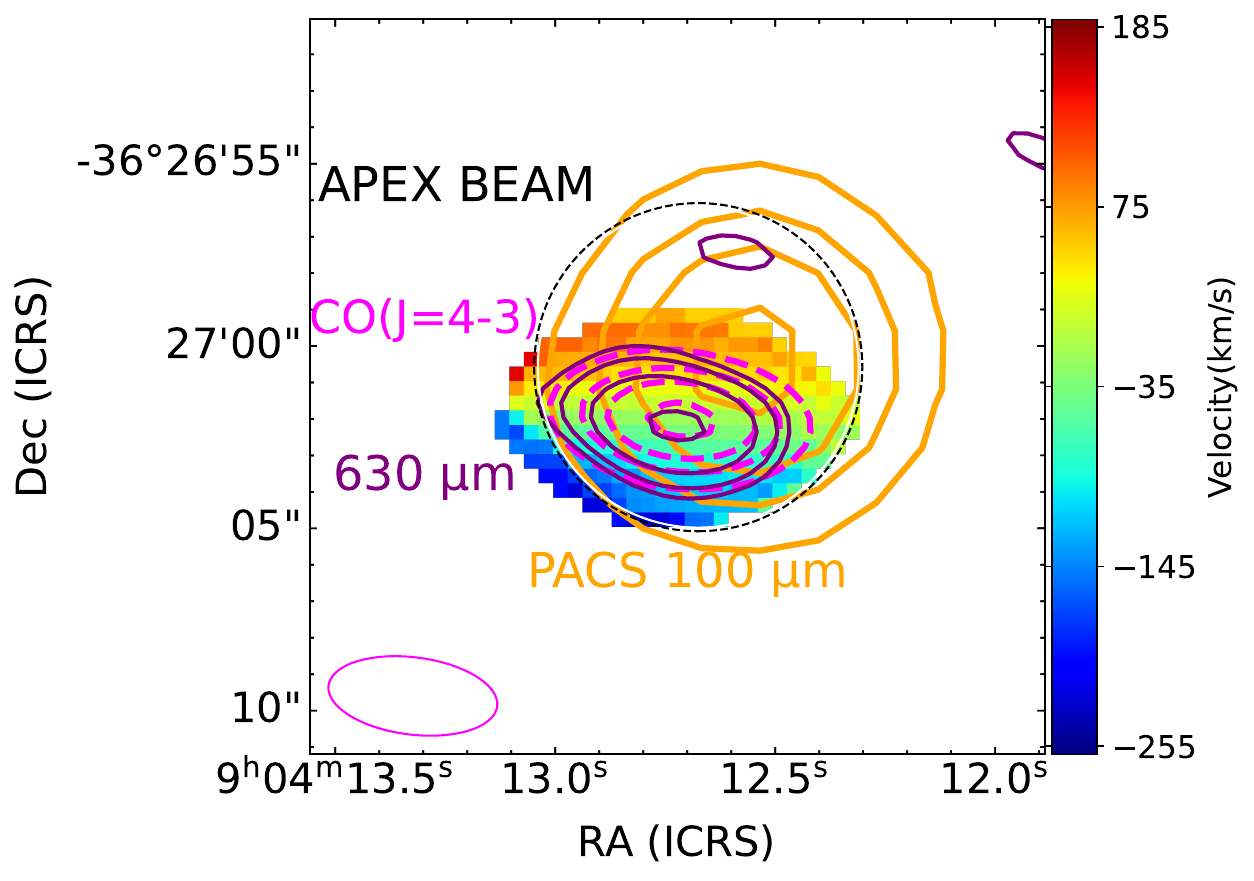}

    \caption{ Comparison between the CO(3--2), CO(4--3) line and the p-\htot211202 for the ULIRGs in our sample. {\it Left}: Overplotted APEX p-\htot211202 and ALMA CO(3--2), CO(4--3) spectra (smoothed to the resolution of the APEX beam before extraction). Both spectra are arbitrarily normalized to the peak of a simple Gaussian fit (red for p-\htot211202 and blue for CO(4--3)) in order to compare the profiles.  In the lower panel, we show the residuals between the profiles and the Gaussian fits. {\it Right}: Velocity field (moment 1) of CO($J$\,=\,3--2), CO($J$\,=\,4--3) (colorbar) compared to the PACS 100\,$\mu$m dust emission(orange contours) and ALMA intensity CO($J$\,=\,4--3) (dashed magenta) contours, 630\,$\mu$m (purple contours) and 850\,$\mu$m (purple contours). 
    {\color{black}The synthesised beams of ALMA CO emission are shown at the bottom left of the right hand side column.}}
    \label{fig:co43}
\end{figure*}

\subsection{Spatial and spectral comparison between H$_2$O, CO, HCN and dust}

In order to get insights {\color{black}on the origin of the excitation mechanisms} of the p-\htot211202\ emission, in Figure~\ref{fig:BEAM}, we present the APEX {\color{black}pointing overplotted on the ALMA} CO($J$\,=\,3--2), CO($J$\,=\,4--3) and HCN($J$\,=\,2--1) integrated line contours. 
{\color{black}Because 101$\mu$m photons are predominantly exciting the \hto\ 2$_{20}$ level that relates to the p-\htot211202\ line \citep[e.g.][]{Liu2017, Gonzales2014}, we explore the spatial distribution of the 100\,$\mu$m dust continuum using the PACS-100$\mu$m images in relation to that of the \hto\ line.} We note that the PACS 100\,$\mu$m emission in IRAS\,17208-0014 and IRAS\,09022-3615 is point-like (at $9\farcs5$ {\sc fwhm} $\sim$ 14 kpc); hence most of the far-IR photons come from the innermost central regions of these ULIRGs. The interferometric ALMA observations toward IRAS\,06035-7102 show the HCN($J$\,=\,2--1) emission has a deconvolved major axis FWHM of $0.71\pm0.13$\,kpc associated with a central region, while the CO($J$\,=\,3--2) is spatially aligned with HCN($J$\,=\,2--1) presenting a deconvolved major axis FWHM size of $3.1\pm1.0$\,kpc. In IRAS\,17208-0014, the APEX pointing is centred on the CO($J$\,=\,4--3) emission, and {\color{black}the PACS-100\,$\mu$m emission is centred on the same region}. However, in IRAS\,09022-3615, we see a significant offset between the peak position of the PACS-100\,$\mu$m photometry and the CO($J$\,=\,4--3) {\color{black}emission (Figures~\ref{fig:BEAM}, \ref{fig:co43} $\&$ \ref{fig:dust}), while at the same time, the APEX pointing is slightly offset with respect to their peaks, located in the redshifted part of the CO emission.}



The left panel of Figure~\ref{fig:co43} shows a normalised spectroscopic comparison between the p-\htot211202, CO($J$\,=\,3--2) and CO($J$\,=\,4--3) 
at 50\,km\,s$^{-1}$ channel width. The velocities from ALMA and APEX spectra are under the same frame LSRK (Doppler correction was already applied during the observations). In  IRAS\,06035-7102, {\color{black}both CO($J$\,=\,3--2) and the p-\htot211202} emission are spectroscopically aligned. The same is seen for IRAS\,17208-0014 when comparing CO($J$\,=\,4--3) and p-\htot211202. Nevertheless, in IRAS\,09022-3615, we find that the water emission is redshifted compared to the gas traced by CO($J$\,=\,4--3). This is best illustrated by the Gaussian fits of the p-\htot211202 (red) and CO($J$\,=\,4--3) (blue), where we find that \hto\ has a peak emission at about $+100$\,km\,s$^{-1}$ with respect to the systemic velocity defined by the redshift of the sources (see Table~\ref{tab:integrated_flux}). This is evident from the residual spectrum, where we can see that there is an excess of water emission in the red part of the spectra at a similar velocity while having a negative peak at the blue part. 

The right panel of Figure~\ref{fig:co43} shows for IRAS\,06035-7102 the intensity and moment 1 map of the CO($J$\,=\,3--2) together with the contours of the intensity of dust emission at 850\,$\mu$m. We show that the APEX pointing is aligned with the 850\,$\mu$m ALMA dust continuum and with the CO($J$\,=\,3--2), suggesting that all the different signals come from the same compact region. For IRAS\,17208-0014 and IRAS\,09022-3615, we present the intensity and moment 1 map of the CO($J$\,=\,4--3) with respect to the contours of the intensity of dust emission of PACS\,100\,$\mu$m and ALMA\,630\,$\mu$m. In IRAS\,17208-0014, the CO($J$\,=\,4--3), the 100\,$\mu$m and 630\,$\mu$m dust continuum {\color{black}are well aligned each other}. This suggests that the different tracers are revealing the same dense gas from the very central part of the galaxy. In IRAS\,09022-3615, both CO($J$\,=\,4--3) and the 630\,$\mu$m dust continuum are spatially aligned, nevertheless the PACS-100\,$\mu$m emission is spatially offset by $\sim2\farcs7$ (2.9\,kpc).
{\color{black} If such a spatial offset is real, we might be finding an explanation for the $+100$\,km\,s$^{-1}$ offset between the p-\htot211202 and CO(4--3) lines, where possibly the water line is excited by infrared pumping, via the $101$\,$\mu$m pumping path, at a region different than where the CO line is produced, as the PACS-100\,$\mu$m emission is peaking at the redshifted part of the CO emission. In this case, the offset between $100$\,$\mu$m (PACS) and 630\,$\mu$m (ALMA) might be due to temperature/optical-depth variation across the galaxy.}
After exploring photometric data taken with the {\it Spitzer Space Observatory} at 8$\mu$m and 24$\mu$m, we confirm a good alignment between ALMA\,630\,$\mu$m and {\it Spitzer} data. {\color{black}Considering that PACS photometry lies between these two wavelengths, this suggests instead that the observed offset might be attributed to issues with the astrometric calibration of the {\it Herschel}-PACS observation. The absolute pointing {\it Herschel} error in scan maps can range from $0\farcs9$ to $2\farcs36$ (see Table 5 of \citealt{Pacs2014}),
an uncertainty which is not consistent but varies at each specific observing epoch subject to unique conditions. Our observed spatial offset is approximately the upper limit of the astrometric error.
If the offset is indeed caused by the astrometric error, nevertheless, the origin of the spectroscopic offset remains unclear. However, in the literature, high spectral resolution observations have indicated the presence of variations in the line profiles of CO($J$\,=\,3--2) and p-\htot211202\ in Mrk 232 and NGC 6240 (as depicted in Figure 2 from \citealt{Liu2017}), nevertheless these differences have not been explored in detail. It is worth noting that IRAS\,09022-3615 has a complex morphology, as shown in Figure~\ref{fig:dust}, where the peak of MIPS 24\,$\mu$m and ALMA 630\,$\mu$m is not aligned with the R-band {\it HST} image.  }


Due to the limited spatial resolution of APEX, to fully understand the origin and powering source of the submm H$_2$O emission, higher angular resolution observations, such as ALMA Band-9 observations, are needed. These resolved observations need to be integrated into detailed photo-dissociation and radiative transfer modelling to tackle the nature of the p-\htot211202\ emission and provide insights on the usage of this line to describe the warm ISM in galaxies.

{\color{black}}


\section{Conclusion}

We present the first ground-based detection {\color{black}of the p-\htot211202 emission line at 752.033\,GHz in local ULIRGs using
APEX SEPIA Band 9 towards} 
IRAS\,06035-7102 ($z$\,=\,0.07946), IRAS\,17208-0014 ($z$\,=\,0.04281) and IRAS\,09022-3615 ($z$\,=\,0.05964).
We demonstrate that despite the low transmission of
the atmosphere at these high frequencies, {\color{black}observing this
thermal water line in local galaxies can be done using ground-based facilities operating at high frequencies, reaching
high signal-to-noise} in a reasonable integration
time ($\sim$ 5-15 hours per source) {\color{black} with high-spectral resolutions. For all three sources, the final APEX spectra are of high quality, improving the previous measurements done by {\it Herschel} SPIRE-FTS both in signal-to-noise ratios and spectral resolution, resulting in the first clear 
detection of this water line in IRAS\,06035-7102.} 

We have spectrally resolved the p-\htot211202 line and derived their velocity-integrated flux densities and intrinsic luminosities. We find that the H$_2$O emission follows the \lhto - \ir correlation.
We compared the p-\htot211202 line emission to the dust continuum at different wavelengths 100\,$\mu$m, 630\,$\mu$m, 850\,$\mu$m and the molecular gas traced by CO($J$\,=\,3--2), CO($J$\,=\,4--3) and HCN($J$\,=\,2--1). 
{\color{black}In IRAS\,06035-7102, the water emission is co-spatial with the dust emission at 850\,$\mu$m and spectrally} aligned with the gas traced by CO($J$\,=\,3--2). 
In IRAS\,17208-0014, a similar behaviour is seen, where the continuum dust emission at 100\,$\mu$m and 630\,$\mu$m are spatially aligned, while in frequency space the gas traced by CO($J$\,=\,4--3) is spectrally aligned with the p-\htot211202 line.
{\color{black} In the case of IRAS\,09022-3615, the p-\htot211202\ water emission line is redshifted by 100 km\,s$^{-1}$ as compared with the CO($J$\,=\,4--3) emission line. The origin of this offset is still inconclusive, and higher angular resolution observations are needed to understand the reasons behind this shift.}

Our pilot APEX survey opens a new window {\color{black}for studying submm \hto\ emission lines in local ULIRGs using ground-based facilities,  offering better sensitivity and spectral resolution than the past space-based facilities, such as {\it Herschel}. Future higher spatial resolution observations,  using ALMA Band-9, together} with detailed radiative transfer modelling, can help us explore the distribution of the molecular gas and {\color{black} dust continuum in local ULIRGs to disentangle the origin of the exciting mechanism of the p-\htot211202 line by reaching down to the scales of giant molecular clouds.}

\section*{Acknowledgements}
{\color{black}We are thankful to the referee, Pierre Cox, for his valuable suggestions, which have significantly enhanced the quality of this paper.} DQ acknowledges support from the National Agency for Research and Development (ANID)/Scholarship Program/Doctorado Nacional/2021-21212222. 
 C.Y. and S.A. acknowledge support from ERC Advanced Grant 789410.
 E.I. acknowledges funding by ANID FONDECYT Regular 1221846. {\color{black}EG-A thanks the Spanish MICINN for support under projects PID2019-105552RB-C41 and PID2022-137779OB-C41}. Y.J. acknowledges financial support from ANID BASAL project No. FB210003.
This publication is based on data acquired with the Atacama Pathfinder Experiment (APEX), project ID 103.B-0471. APEX is a collaboration between the Max-Planck-Institut fur Radioastronomie, the European Southern Observatory, and the Onsala Space Observatory. This publication makes use of the following ALMA data: \url{ADS/JAO.ALMA#2017.1.00022.S}, \url{ADS/JAO.ALMA#2018.1.00994.S} ALMA is a partnership of ESO (representing its member states), NSF (USA) and NINS (Japan), together with NRC (Canada), MOST and ASIAA (Taiwan), and KASI (Republic of Korea), in cooperation with the Republic of Chile.
The Joint ALMA Observatory is operated by ESO, AUI/NRAO and NAOJ. PACS has been developed by a consortium of institutes led by MPE (Germany) and including UVIE (Austria); KU Leuven, CSL, IMEC (Belgium); CEA, LAM (France); MPIA (Germany); INAF-IFSI/OAA/OAP/OAT, LENS, SISSA (Italy); IAC (Spain). This development has been supported by the funding agencies BMVIT (Austria), ESA-PRODEX (Belgium), CEA/CNES (France), DLR (Germany), ASI/INAF (Italy), and CICYT/MCYT (Spain). SPIRE has been developed by a consortium of institutes led by Cardiff University (UK) and including Univ. Lethbridge (Canada); NAOC (China); CEA, LAM (France); IFSI, Univ. Padua (Italy); IAC (Spain); Stockholm Observatory (Sweden); Imperial College London, RAL, UCL-MSSL, UKATC, Univ. Sussex (UK); and Caltech, JPL, NHSC, Univ. Colorado (USA). This development has been supported by national funding agencies: CSA (Canada); NAOC (China); CEA, CNES, CNRS (France); ASI (Italy); MCINN (Spain); SNSB (Sweden); STFC, UKSA (UK); and NASA (USA). This work is based [in part] on observations made with the Spitzer Space Telescope, which is operated by the Jet Propulsion Laboratory, California Institute of Technology under a contract with NASA.

\section*{Data Availability}

The data underlying this article are available in the ESO Science
Archive at https://archive.eso.org/scienceportal/home, and can be accessed
with the Project ID: 103.B-0471



\bibliographystyle{mnras}
\bibliography{example} 




\clearpage

\appendix

\section{Photometry of the galaxies}
The following Table~\ref{tab:photometry} presents the photometric fluxes derived from the PACS analyses at 100 $\mu$m and 160 $\mu$m (see Section~\ref{sed_section}). Data is presented together with fluxes taken from the literature.
\bigskip
 

\begin{table*}
	\caption{Multi-wavelength infrared photometry for the three galaxies presented in Table~\ref{tab:lineparameters}. Flux densities are shown in Jy units.}
	\label{tab:photometry}
	\begin{tabular}[hbpt]{lccccccccr} 
	\hline
	N   &WISE     &IRAS     &PACS      &IRAS      &PACS   &PACS   &SPIRE     &SPIRE     &SPIRE \\
	      &22$\mu$m &60$\mu$m & 70$\mu$m&100$\mu$m &100$\mu$m &160$\mu$m &250$\mu$m &350$\mu$m &500$\mu$m \\
	&[Jy]&[Jy]&[Jy]&[Jy]&[Jy]&[Jy]&[Jy]&[Jy]&[Jy]\\
		\hline
	1&	0.214	$\pm$	0.001	&	5.13	$\pm$	0.15	&	&5.65	$\pm$	0.28	&				&				&	1.23	$\pm$	0.02	&	0.397	$\pm$	0.001	&	0.130	$\pm$	0.008	\\
 	2&	0.548	$\pm$	0.003	&	31.14	$\pm$	1.87	&	38.09	$\pm$	1.74	&34.90	$\pm$	2.09	&	38.58	$\pm$	0.02	&	23.31	$\pm$	0.10	&	7.92	$\pm$	0.04	&	2.95	$\pm$	0.01	&	0.954	$\pm$	0.009	\\
	3&	0.430	$\pm$	0.002	&	11.47	$\pm$	0.57	&	12.81 $\pm$ 0.58&			&	13.25	$\pm$	0.02	&	8.09	$\pm$	0.08	&	2.449	$\pm$	0.007	&	0.823	$\pm$	0.004	&	0.252	$\pm$	0.005	\\
		\hline
	\end{tabular}\\
	\textbf{Note:} WISE emission at 22\,$\mu$m from AllWISE Source Catalog \citep[][]{AIIWISE2014}. IRAS 60 and 100\,$\mu$m measurements again retrieved from NASA/IPAC Extragalactic Database(NED). PACS emission at 70\,$\mu$m from \citet{Chu2017}. SPIRE emission at 250\,$\mu$m, 350\,$\mu$m and 500\,$\mu$m from \citet{Clements2018}
	
\end{table*}
\section{{\it Herschel} SPIRE-FTS}
The following Figure~\ref{fig:SPIRE} presents the full spectra obtained by {\it Herschel} SPIRE-FTS for IRAS6035-7102, IRAS17208-0014 and IRAS09022-3615

\begin{figure*}
    \includegraphics[scale=0.38]{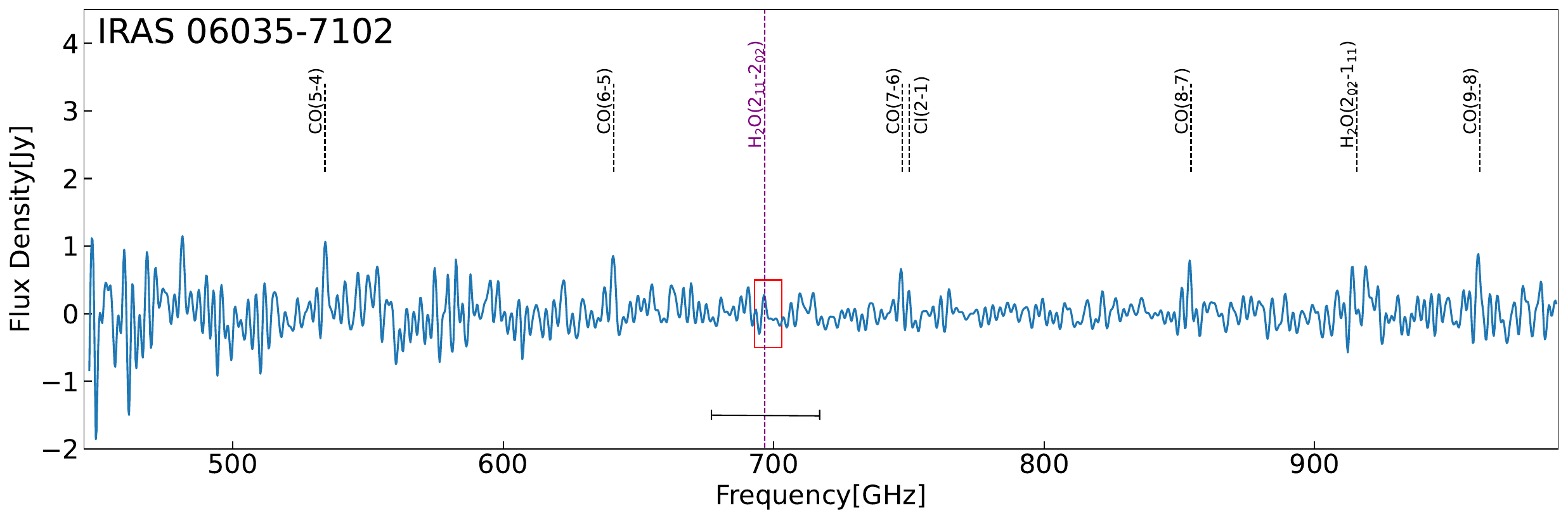}
	\includegraphics[scale=0.38]{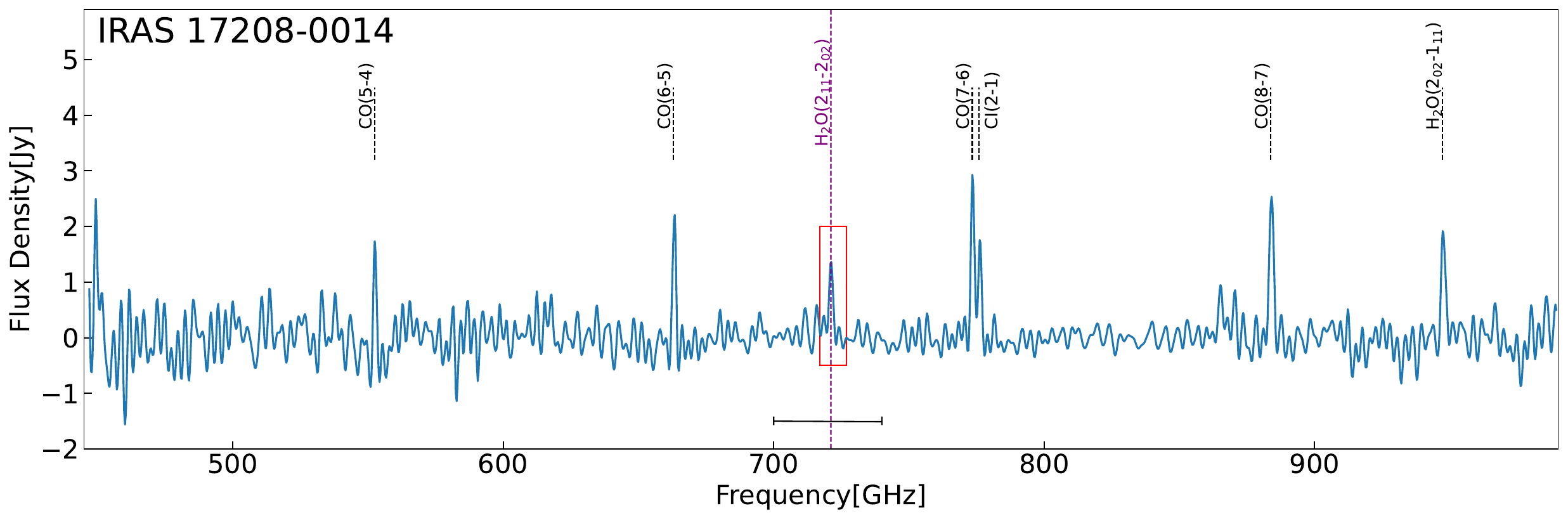} 
    \includegraphics[scale=0.38]{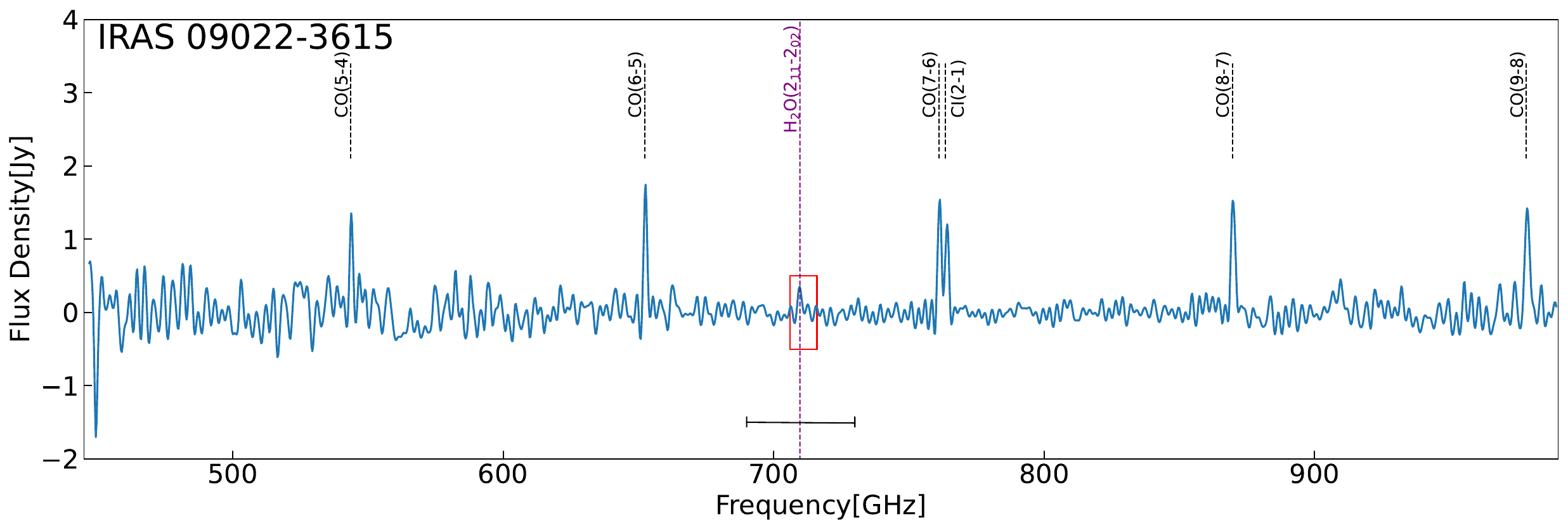}

    \caption{{\it Herschel} SPIRE-FTS spectra for the three sources. Black segments indicate the bandwidth where we measured the RMS. Red squares show the spectra taken to compare with our observations. The dashed magenta line is  p-\htot211202. In IRAS\, 6035-7102, the water line was not detected with {\it Herschel} SPIRE-FTS }
    \label{fig:SPIRE}
\end{figure*}

\section{Dust continuum}
The following Figure~\ref{fig:dust} compares different dust continuum observations towards IRAS 09022-3615. A comparison between the emission PACS\,100\,$\mu$m, Spitzer-MIPS\,24\,$\mu$m and ALMA\,630\,$\mu$m are shown. {\color{black} Considering the alignment between the {\it Spitzer}-24\,$\mu$m and ALMA-630\,$\mu$m photometry, and that PACS photometry lies between these two wavelengths, the figure suggests the $2\farcs9$ PACS offset (see Figure~\ref{fig:co43}) might be related to a calibration issue with the {\it Herschel} astrometry.}

\begin{figure*}
    \includegraphics[scale=0.38]{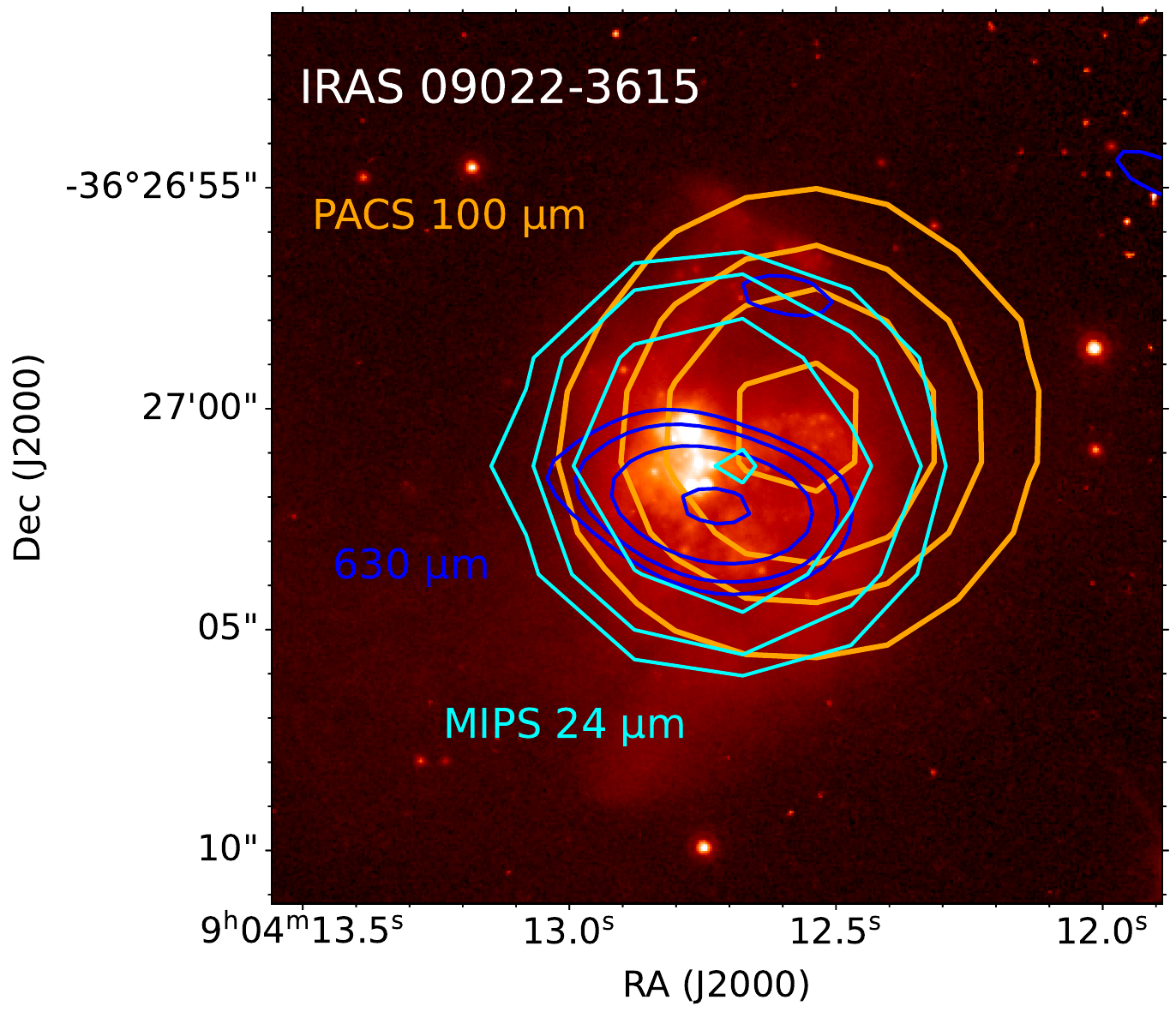}

    \caption{Different dust continuum emissions in IRAS\, 09022-3615 shown on top of a {\it Hubble Space Telescope} R-band image. The contours represent the 20$\%$, 40$\%$, 60$\%$, 90 $\%$ of the maximum peak intensity at 100\,$\mu$m from PACS (orange) and 630\,$\mu$m from ALMA (blue) and Spitzer-MIPS\,24$\mu$m (cyan).}
    \label{fig:dust}
\end{figure*}
\bsp	
\label{lastpage}
\end{document}